\documentclass{article}
\usepackage[utf8]{inputenc}
\usepackage{fullpage}
\usepackage{graphicx}
\usepackage{comment}
\usepackage{float}

\usepackage{titling}

\begin{document}

\title{A Brief Study of Open Source Graph Databases}

\author{Rob McColl~~David Ediger~~Jason Poovey~~Dan Campbell~~David Bader\\Georgia Tech Research Institute, Georgia Institute of Technology}
\date{}

\maketitle

\begin{abstract}
With the proliferation of large irregular sparse relational datasets, new storage and
analysis platforms have arisen to fill gaps in performance and capability left by
conventional approaches built on traditional database technologies and 
query languages.  Many of these platforms apply graph structures and analysis techniques 
to enable users to ingest, update, query and compute on the topological structure of these 
relationships represented as set(s) of edges between set(s) of vertices.  To store and 
process Facebook-scale datasets, they must be able to support data sources with 
billions of edges, update rates of millions of updates per 
second, and complex analysis kernels.  These platforms must provide intuitive 
interfaces that enable graph experts and novice programmers to write implementations 
of common graph algorithms.

In this paper, we explore a variety of graph analysis and storage platforms.  We compare 
their capabilities, interfaces, and performance by implementing and computing a set 
of real-world graph algorithms on synthetic graphs with up to 256 million edges.
In the spirit of full disclosure,
several authors are affiliated with the development of STINGER.
\end{abstract}

\section{Background}
In the context of this paper, the term \emph{graph database} is used
to refer to any storage system that can contain, represent, and query a 
graph consisting of a set of vertices and a set of edges relating pairs of vertices.  
This broad definition encompasses many technologies.  Specifically, we examine schemas 
within traditional disk-backed, ACID-compliant SQL databases (SQLite, MySQL, Oracle, and 
MS SQL Server), modern NoSQL databases and graph databases (Neo4j, OrientDB, InfoGrid, 
Titan, FlockDB, ArangoDB, InfiniteGraph, AllegroGraph, DEX, GraphBase, and HyperGraphDB), 
distributed graph processing toolkits based on MapReduce, HDFS, and custom BSP engines 
(Bagel, Hama, Giraph, PEGASUS, Faunus), and in-memory graph packages designed for 
massive shared-memory (NetworkX, Gephi, MTGL, Boost, uRika, and STINGER).  For all of 
these, we build a matrix containing the package maintainer, license, platform, 
implementation language(s), features, cost, transactional 
capabilities, memory vs. disk storage, single-node vs. distributed, text-based
query language support, built-in algorithm support, and primary traversal and query
styles supported.  An abridged version of this matrix is presented in Appendix \ref{app:1}.  
For a selected group of primarily open source graph databases, we implemented and tested
a series of representative common graph kernels on each technology in the set on
the same set of synthetic graphs modeled on social networks.

Previous work has compared Neo4j and MySQL using simple queries and breadth-first
search~\cite{neo4j-mysql}, and we are inspired by a combination objective/subjective approach
to the report.  Graph primitives for RDF query languages were extensively studied
in~\cite{RDFprimitives} and data models for graph databases in~\cite{graphdbModels},
which are beyond the scope of this study.  

\subsection*{Algorithms and Approach}

Four common graph kernels were selected with requirements placed on how
each must be implemented to emphasize common implementation
styles of graph algorithms.
The first is the Single Source Shortest Paths (SSSP) problem, which was 
specifically required to be implemented as a level-synchronous parallel breadth-first 
traversal of the graph.  The second is an implementation of the Shiloach-Vishkin 
connected components algorithm (SV)~\cite{SV82} in the edge-parallel label-pushing style.  
The third is PageRank (PR)~\cite{PageRank} in the vertex-parallel Bulk Synchronous Parallel 
(BSP) power-iteration style.  The last is a series of edge insertions and deletions in
parallel to represent random access and modification of the structure.  When it was not
possible to meet these requirements due to restrictions of the framework, the algorithms were
implemented in as close of a manner as possible.  When the framework included an
implementation of any of these algorithms, that implementation was used if it met the 
requirements.  The implementations were intentionally written in a straightforward manner
with no manual tuning or optimization to emulate non-hero programmers.

Many real-world networks demonstrate ``scale-free'' characteristics~\cite{barabasi, WS98}.
Four initial sparse static graphs and sets of edge insertions 
and deletions were created using an R-MAT~\cite{CZF04} generator with parameters $A = 0.55$, $B = 0.1$, 
$C = 0.1$, and $D = 0.25)$.  These graphs contain 1K (tiny), 32K (small), 1M (medium), 
and 16M (large) vertices with 8K, 256K, 8M, and 256M undirected edges, respectively. 
Graphs over 1B edges were considered, but it was determined that many packages 
could not handle that scale.  Edge updates have a 6.25\% probability of being deletions 
uniformly selected from existing edges where deletions are handled in the default manner 
for the package.  For the tiny and small graphs, 100K updates were performed, and 1M 
updates were performed otherwise.

The quantitative measurements taken were initial graph construction time, total memory 
in use by the program upon completion, update rate in edges per second, and time to 
completion for SSSP, SV, and PR.  Tests for single-node platforms and databases were run 
on a system with four AMD Opteron 6282 SE processors and 256GB of DDR3 RAM.
Distributed system tests
were performed on a cluster of 21 systems with minimum configuration of two Intel Xeon X5660
processors
with 24GB DDR3 RAM connected by QDR Infiniband.  HDFS was run in-memory.  Results are
presented in full in Appendix \ref{app:2}.  Packages for which no result is provided either
ran out of memory or simply did not complete in a comparable time.  The code used for
testing is available at \emph{github.com/robmccoll/graphdb-testing}.



\section{Experiences}
\vspace{-2mm}

\subsection{User Experience}
A common goal of open source software is widespread adoption.  Adoption promotes
community involvement, and an active community is likely to contribute use cases, patches,
and new features.  When a user downloads a package, the goal should be to build
as quickly as possible with minimal input from the user.  A number of packages
tested during the course of this study did not build on the first try.  Often, this was
the result of missing libraries or test modules that do not pass, and was resolved
with help from the forums.  Build problems may reduce the likelihood that users
continue with a project.

Among both commercial and open source software packages, we find a lack of consistency
in approaches to documentation.  It is often difficult to quickly get started.
System requirements and library dependencies should be clear.  Quickstart
guides are appreciated.  We find that most packages lack adequate usage examples.
Common examples illustrate how to create a graph and query the existence of vertices
and edges.  Examples are needed that show how to compute graph algorithms, such as
PageRank or connected components, using the provided primitives.  The ideal example
demonstrates data ingest, a real analysis workflow that reveals knowledge, and the
interface that an application would use to conduct this analysis.

While not unique to graphs, there are a multitude input file formats that are
supported by the software packages in this study.  Formats can be built on XML,
JSON, CSV, or other proprietary binary and plain-text formats.  Each has a trade-off 
between size, descriptiveness, and flexibility.  The number of different formats creates a
challenge for data interchange.
Formats that are edge-based, delimited, and self-describing
can easily be parsed in parallel and translated between each other.


\subsection{Developer Experience}
Object-Oriented Programming (OOP) was introduced to enforce modularity, increase reusability, and
add abstraction to enable generality.  These are important goals; however, OOP
can also inadvertently obfuscate logic and create bloated code.  For
example, considering a basic PageRank computation in Giraph, the function implementing the
algorithm uses approximately 16\% of the 275 lines of code.  The remainder of the code registers
various callbacks and sets up the working environment.  Although the extra code provides
flexibility, it can be argued that much is boilerplate.  The framework should
support the programmer spending as little time as possible writing boilerplate and configuration
so that the majority of code is a clear and concise implementation of the algorithm.

A number of graph databases retain many of the properties of a relational database,
representing the graph edges as a key-value store without a schema.  A variety of query
languages are employed.  Like their relational cousins, these databases are ACID-compliant
and disk-backed.  Some ship with graph algorithms implemented natively.
While these query languages may be comfortable to users coming from a database background,
they are rarely sufficient to implement even the most basic of graph algorithms succinctly. 
The better query languages are closer to full scripting environments, which may
indicate that query languages are not sufficient for graph analysis.

In-memory graph databases focus more on the algorithms and rarely provide a query
language.  Access to the graph is done through algorithms or graph-centric APIs.  This 
affords the user the ability to write nearly any algorithm, but presents a certain
complexity and learning curve.  The visitor pattern in MTGL is a strong example of a
graph-centric API.  The BFS visitor allows the user to register callbacks for newly
visited vertices, newly discovered BFS tree edges, and edges to previously discovered
vertices.  Given that breadth-first traversals are frequently used as building blocks
in common graph algorithms, providing an API that performs this traversal over the graph
in an optimal way can abstract the specifics of the data structure while giving
performance benefit to novice users.  Similar primitives might include component labelings,
matchings, independent sets, and others (many of which are provided by NetworkX).
Additionally, listening mechanisms for dynamic algorithms should be considered in the future.


\subsection{Graph-specific Concerns}
An important measurement of performance is the size of memory consumed by the application
processing a graph of interest.  An appropriate measurement of efficiency is the number
of bytes of memory required to store each edge of the graph.  The metadata associated with
vertices and edges can vary by software package.  The largest test graph contained more
than 128 million edges, or about one gigabyte in 4-byte tuple form.  Graph databases
were run on machines with 256GB of memory.  However, a number of
software packages could not run on the largest test graph.

It is often possible to store large graphs on disk, but the working set size of the
algorithm in the midst of the computation can overwhelm the system.  For example, a
breadth-first search will often contain an iteration in which the size of the frontier
is as many as half of the vertices in the graph.  We find that some graph databases can
store these graphs on disk, but cannot compute on them because the in-memory portion
of the computational is too large.

We find differing semantics regarding fundamental
operations on graphs.  For example, when an edge inserted already exists,
there are several possibilities: 1) do nothing, 2) insert another copy of that edge,
3) increment the edge weight of that edge, and 4) replace the existing edge.  
Graph databases may support one or more of these semantics.  The same can be said for
edge deletion and the existence of self-edges.

A common query access pattern among graph databases is the extraction of a subgraph,
egonet, or neighborhood around a vertex.  While answering some questions, a more flexible
interface allows random access to the vertices and edges, as well as graph traversal.
Graph traversal is a fundamental component to many graph algorithms.  In some cases,
random access to graph edges enables cleaner implementation of an algorithm with
better workload balance or communication properties.

\section{Performance}

\begin{figure*}[t]
\centering
\includegraphics[width=0.9\textwidth]{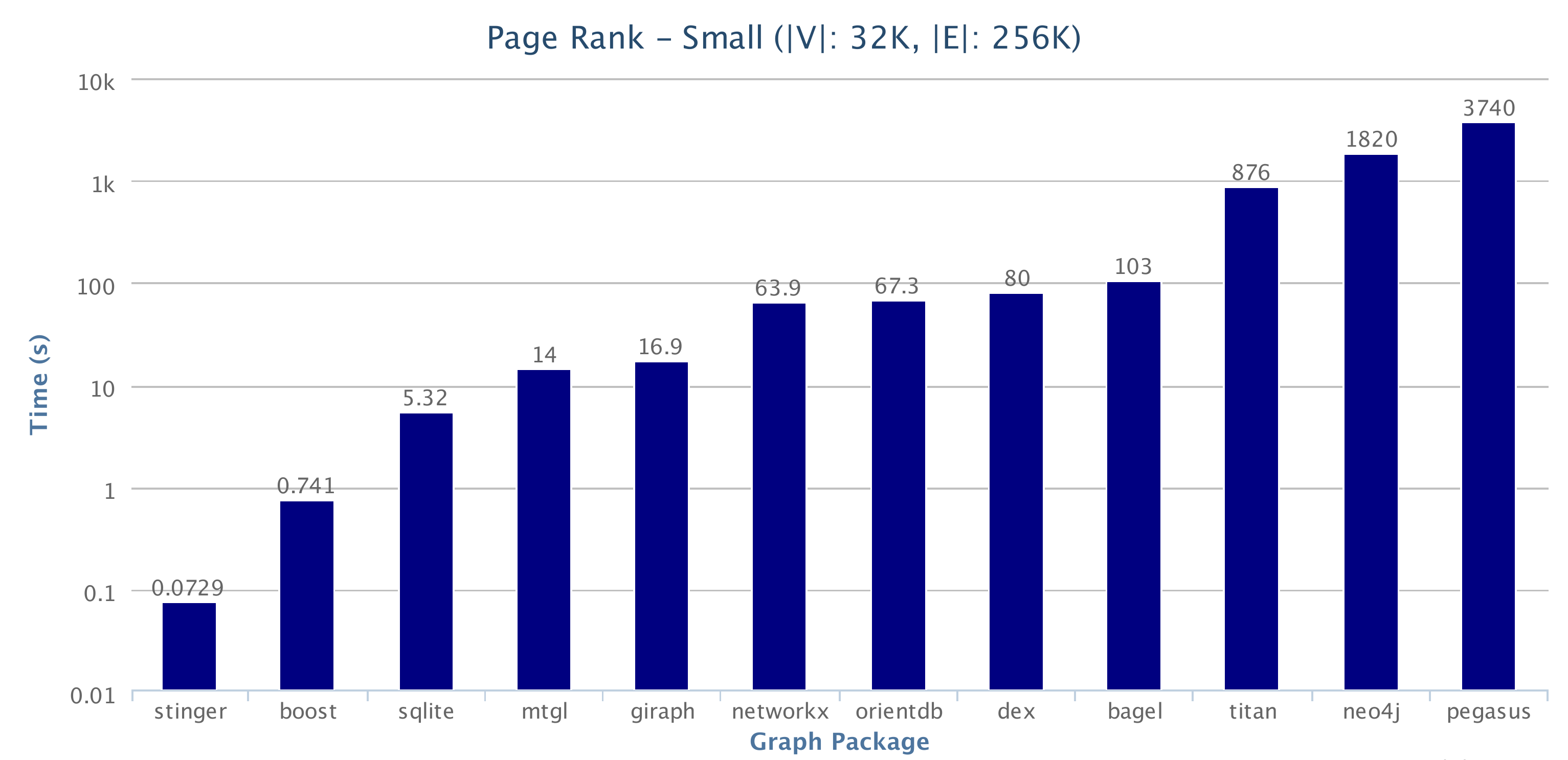}
\caption{The time taken to compute PageRank on the small graph (32K vertices and 256K undirected edges) for each graph analysis package.}
\label{fig:pagerank-small}
\end{figure*}

We compute four algorithms (single source shortest path, connected components, PageRank,
and update) for each of four graphs (tiny, small, medium, and large) using each graph
package.  Developer-provided implementations were used when available.  Otherwise,
implementations were created using the best algorithms, although no heroic programming
or optimization were done.  In general, we find up to five orders of magnitude difference
in execution time.
The time to compute PageRank on a graph with 256K edges is shown in
Figure~\ref{fig:pagerank-small}.  For additional results, please refer to
Appendix~\ref{app:2}.

\section{Position}
We believe there is great interest in graph-based approaches for big data, and many
open source efforts are under way.  The goal of each of these efforts is to extract
knowledge from the data in a more flexible manner than a relational database.
Many approaches to graph databases build upon and leverage the long history of relational
database research.

Our position is that graph databases must become more ``graph aware'' in data
structure and query language.  The ideal graph database must understand analytic queries
that go beyond neighborhood selection.  In relational databases, the index
represents pre-determined knowledge of the structure of the computation without knowing
the specific input parameters.  A relational query selects a subset of the data and
joins it by known fields.  The index accelerates the query by
effectively pre-computing a portion of the query.

In a graph database, the equivalent index is the portion of an analytic or graph algorithm
that can be pre-computed or kept updated regardless of the input parameters of the
query.  Examples include the connected components of the graph, algorithms based on
spanning trees, and vertex statistics like PageRank or clustering coefficients.  A sample
query might ask for shortest paths vertices, or alternatively the
top $k$ vertices in the graph according to PageRank.  Rather than compute the answer
on demand, maintaining the analytic online results in lower latency responses.  While indices
over the properties of vertices may be convenient for certain cases, these types of queries
are served equally well by traditional SQL databases.

While many software applications are backed by databases, most end users are unaware of
the SQL interface between application and data.  SQL is not itself an application.
Likewise, we expect that NoSQL is not itself an application.  Software will be built
atop NoSQL interfaces that will be hidden from the user.  The successful NoSQL framework
will be the one that enables algorithm and application developers to implement their
ideas easily while maintaining high levels of performance.

Visualization frameworks are often incorporated into graph databases.  The result of a
query is returned as a visualization of the extracted subgraph.  State-of-the-art
network visualization techniques rarely scale to more than one thousand vertices.
We believe that relying on visualization for query output limits the types of queries
that can be asked.  Queries based on temporal and semantic changes can be difficult to
capture in a force-based layout.  Alternative strategies include visualizing statistics
of the output of the algorithms over time, rather than the topological structure of the
graph.

With regard to the performance of disk-backed databases, transactional guarantees may
unnecessarily reduce performance.
It could be argued that such guarantees are not necessary in the average case, especially 
atomicity and durability.  For example, if the goal is to capture and analyze a stream
of noisy data from Twitter, it may be acceptable for an edge connecting two users to briefly
appear in one direction and not the other.  Similarly, in the event of a power failure, 
the loss of a few seconds of data may not significantly change which vertices have the
highest centrality on the graph.  Some of the databases presented here seem to have reached
this realization and have made transactional guarantees optional.


\bibliographystyle{plain}
\bibliography{sources}

\newpage

\appendix
\section{Graph Databases}
\label{app:1}

\begin{figure}[h!]
\centering
\includegraphics[angle=270,width=0.6\textwidth]{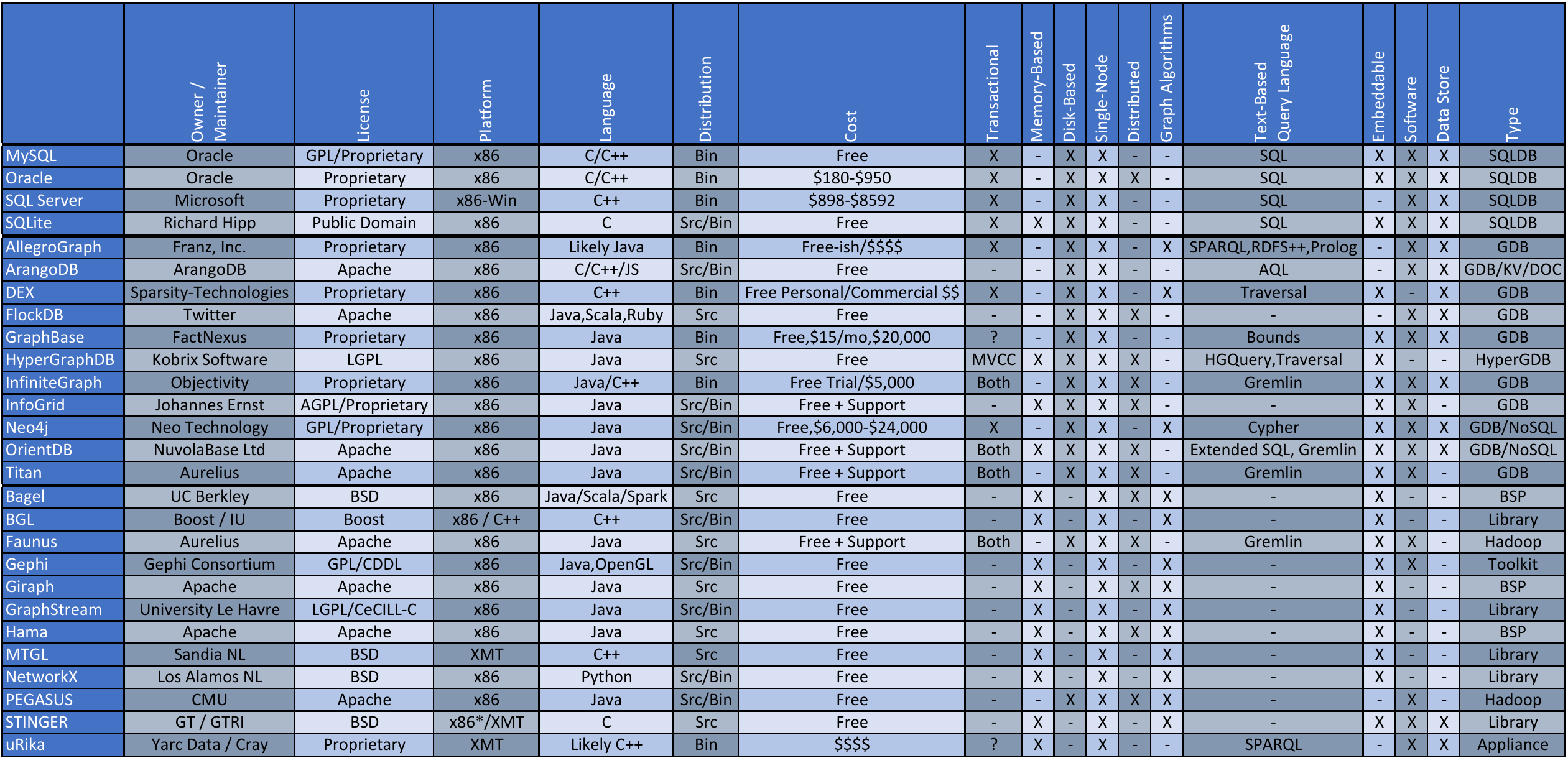}
\label{fig:comparison-matrix}
\end{figure}

\section{Performance Results}
\label{app:2}

\subsection{Tiny Graph}

\begin{figure}[H]
\centering
\includegraphics[width=0.9\textwidth]{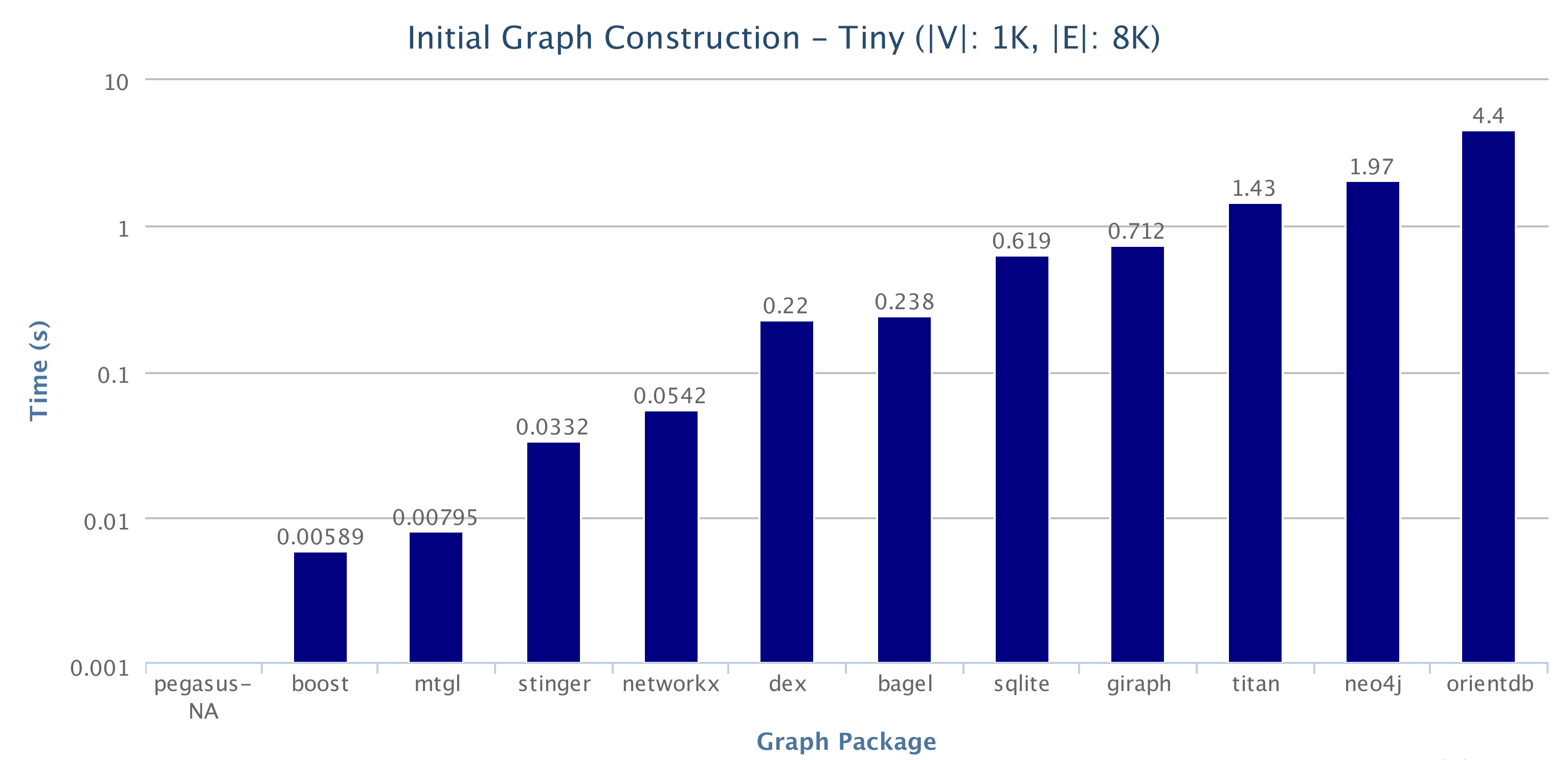}
\caption{The time taken to construct the tiny graph (1K vertices and 8K undirected edges) for each graph analysis package.}
\end{figure}

\begin{figure}[H]
\centering
\includegraphics[width=0.9\textwidth]{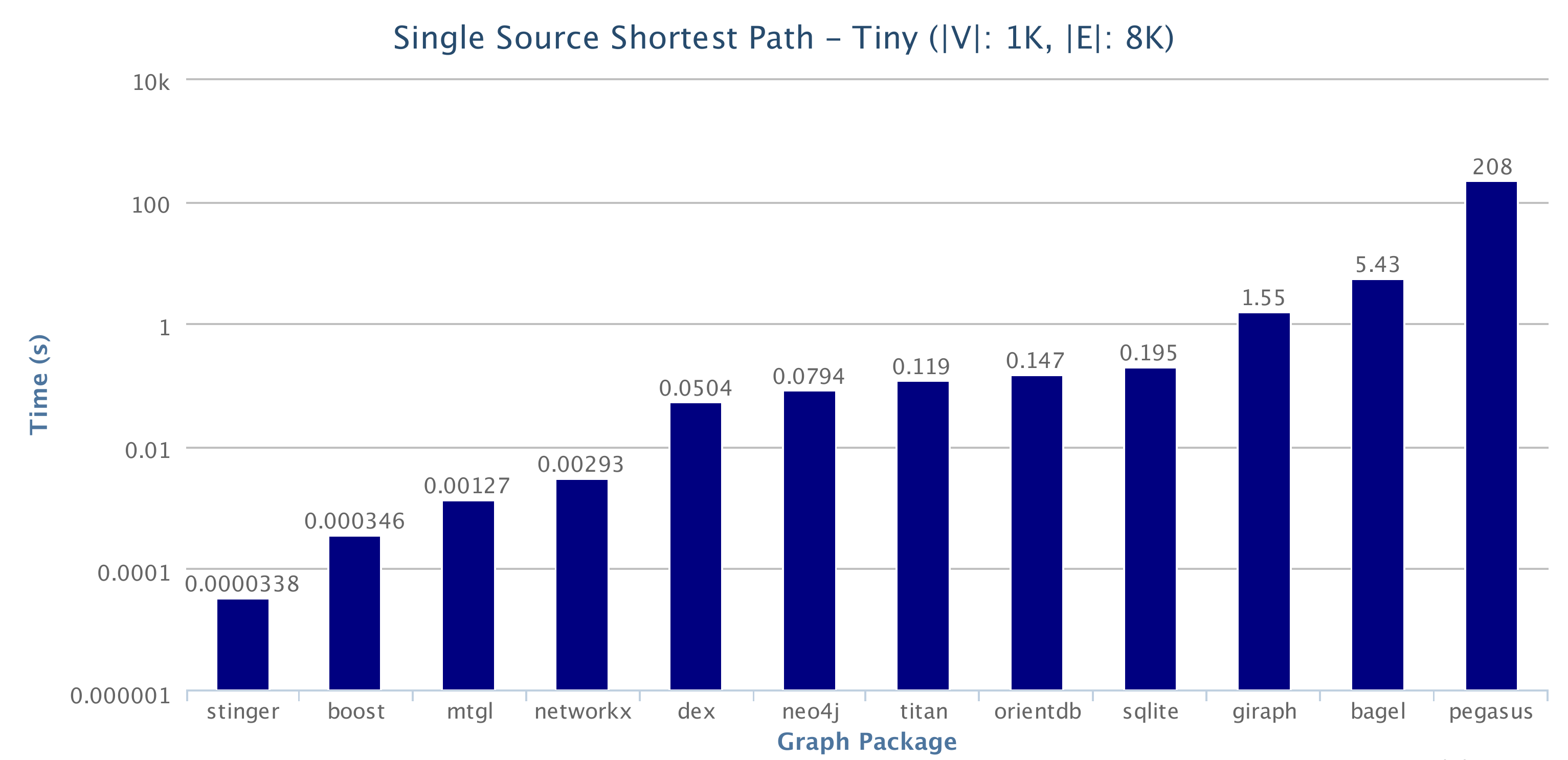}
\caption{The time taken to compute SSSP from vertex 0 in the tiny graph (1K vertices and 8K undirected edges) for each graph analysis package.}
\end{figure}

\begin{figure}[H]
\centering
\includegraphics[width=0.9\textwidth]{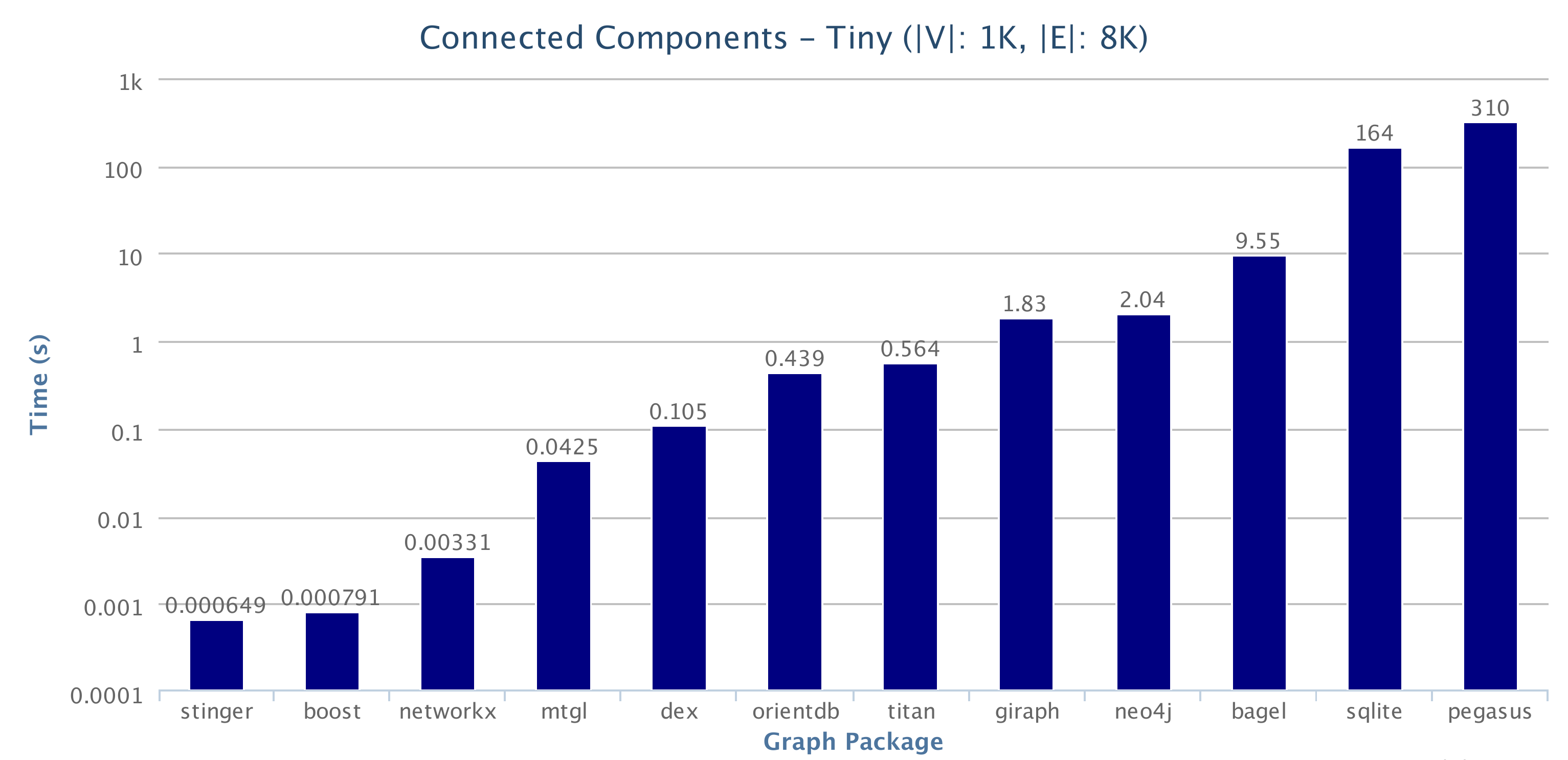}
\caption{The time taken to label the connected components of the tiny graph (1K vertices and 8K undirected edges) for each graph analysis package.}
\end{figure}

\begin{figure}[H]
\centering
\includegraphics[width=0.9\textwidth]{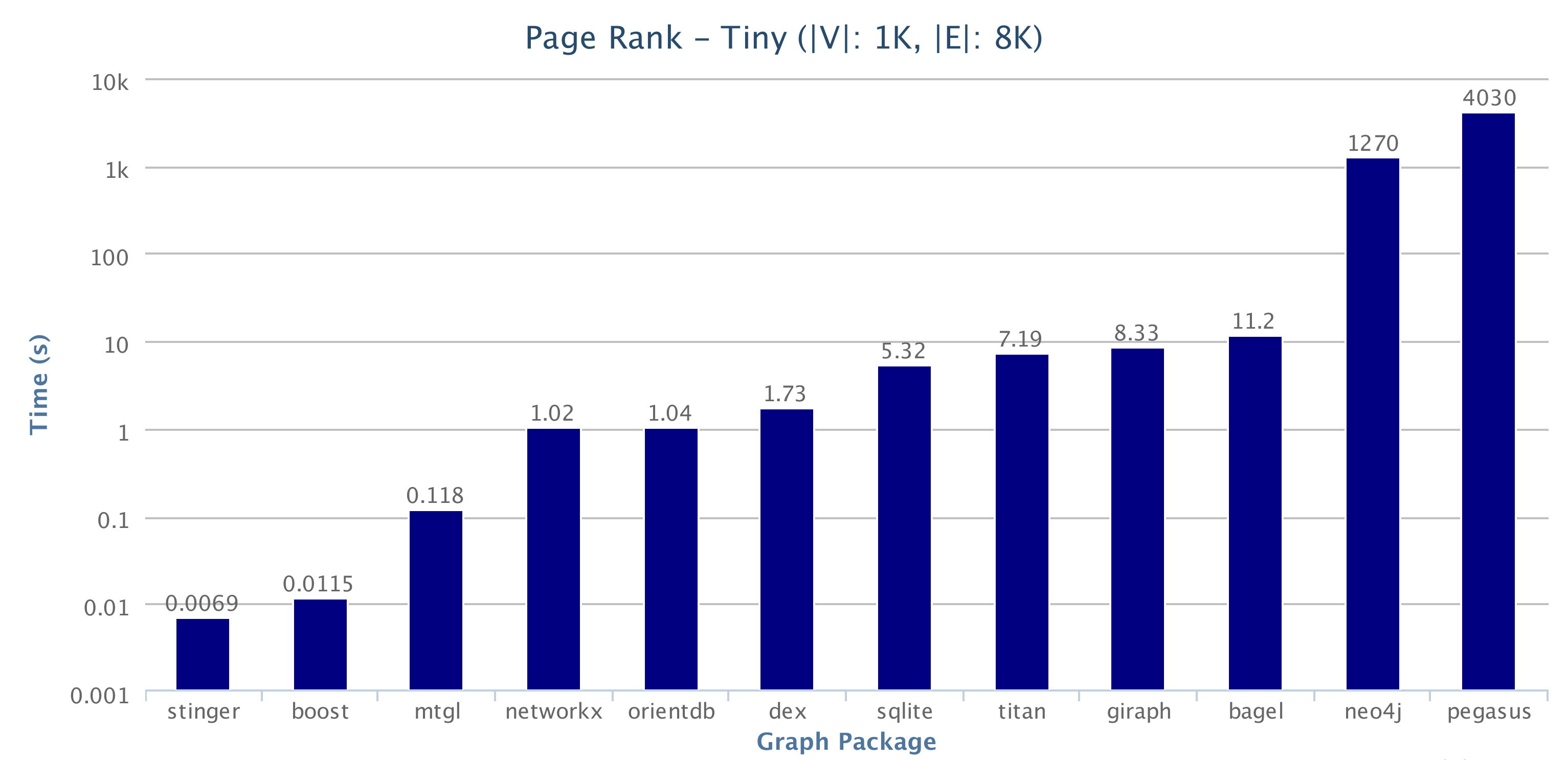}
\caption{The time taken to compute the PageRank of each vertex in the tiny graph (1K vertices and 8K undirected edges) for each graph analysis package.}
\end{figure}

\begin{figure}[H]
\centering
\includegraphics[width=0.9\textwidth]{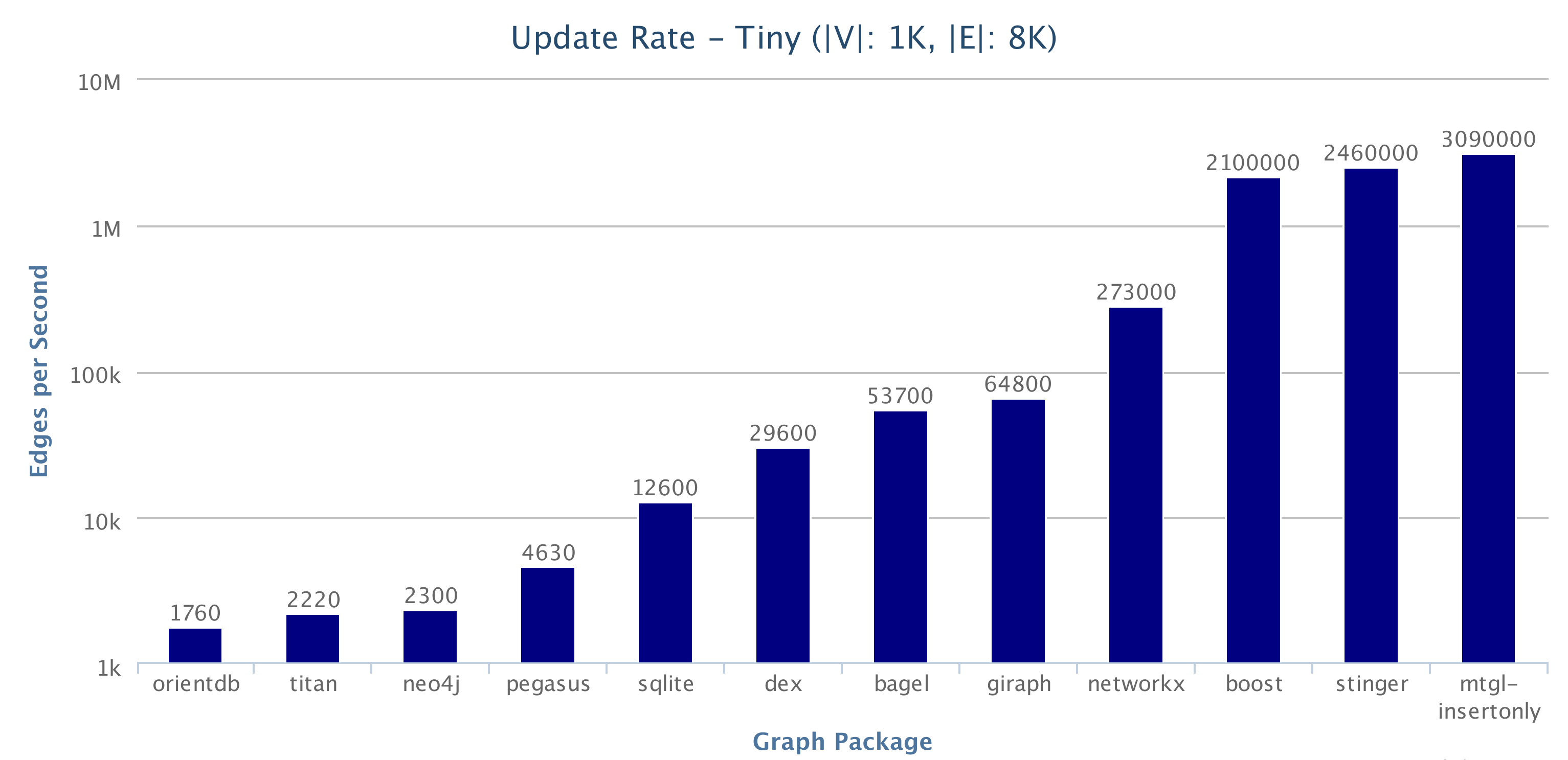}
\caption{The update rate in edges inserted/deleted per second when applying 100,000 edge updates to the tiny graph (1K vertices and 8K undirected edges) for each graph analysis package.}
\end{figure}

\begin{figure}[H]
\centering
\includegraphics[width=0.9\textwidth]{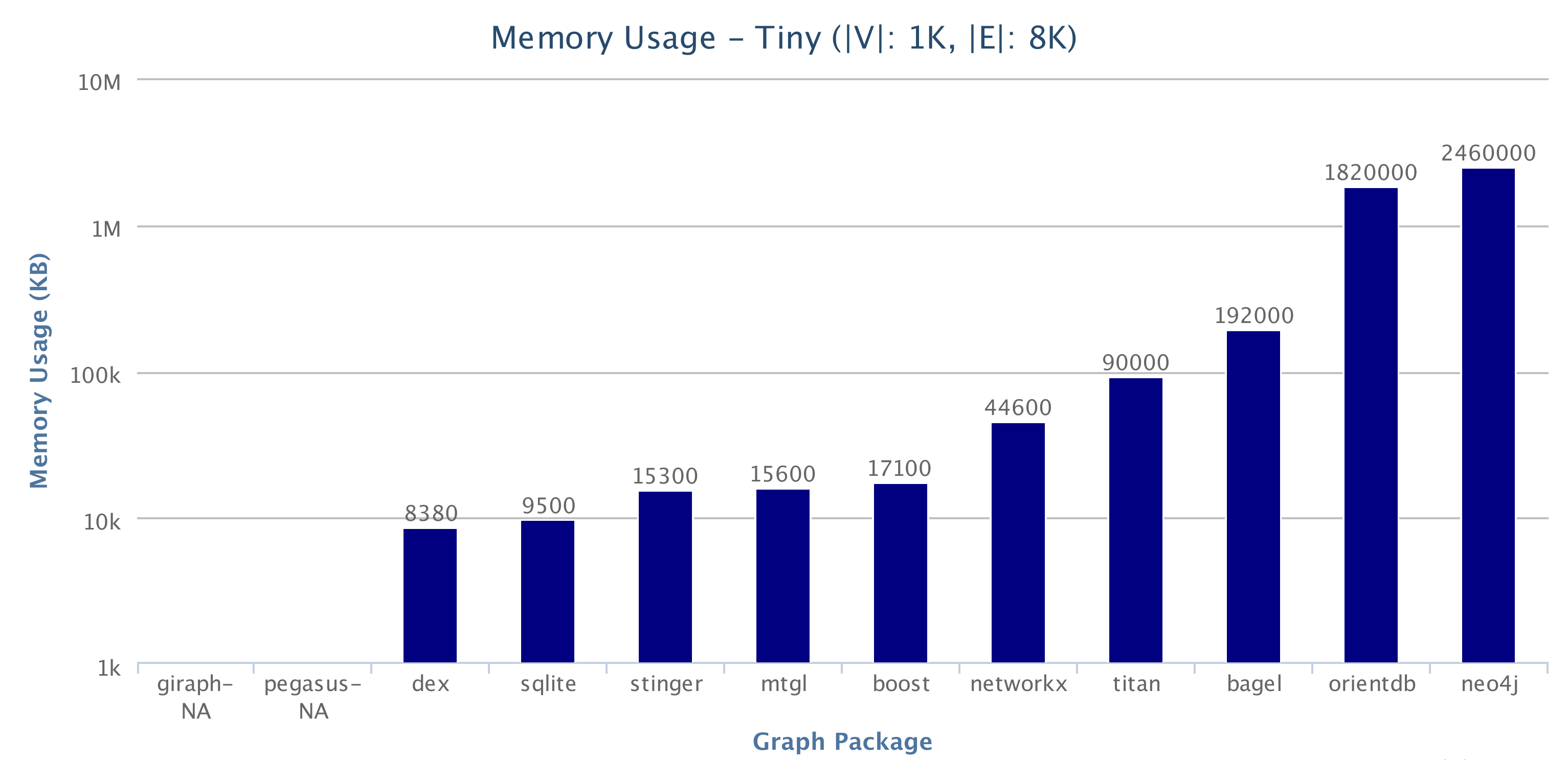}
\caption{The memory occupied by the program after performing all operations on the tiny graph (1K vertices and 8K undirected edges) for each graph analysis package.}
\end{figure}

\subsection{Small Graph}

\begin{figure}[H]
\centering
\includegraphics[width=0.9\textwidth]{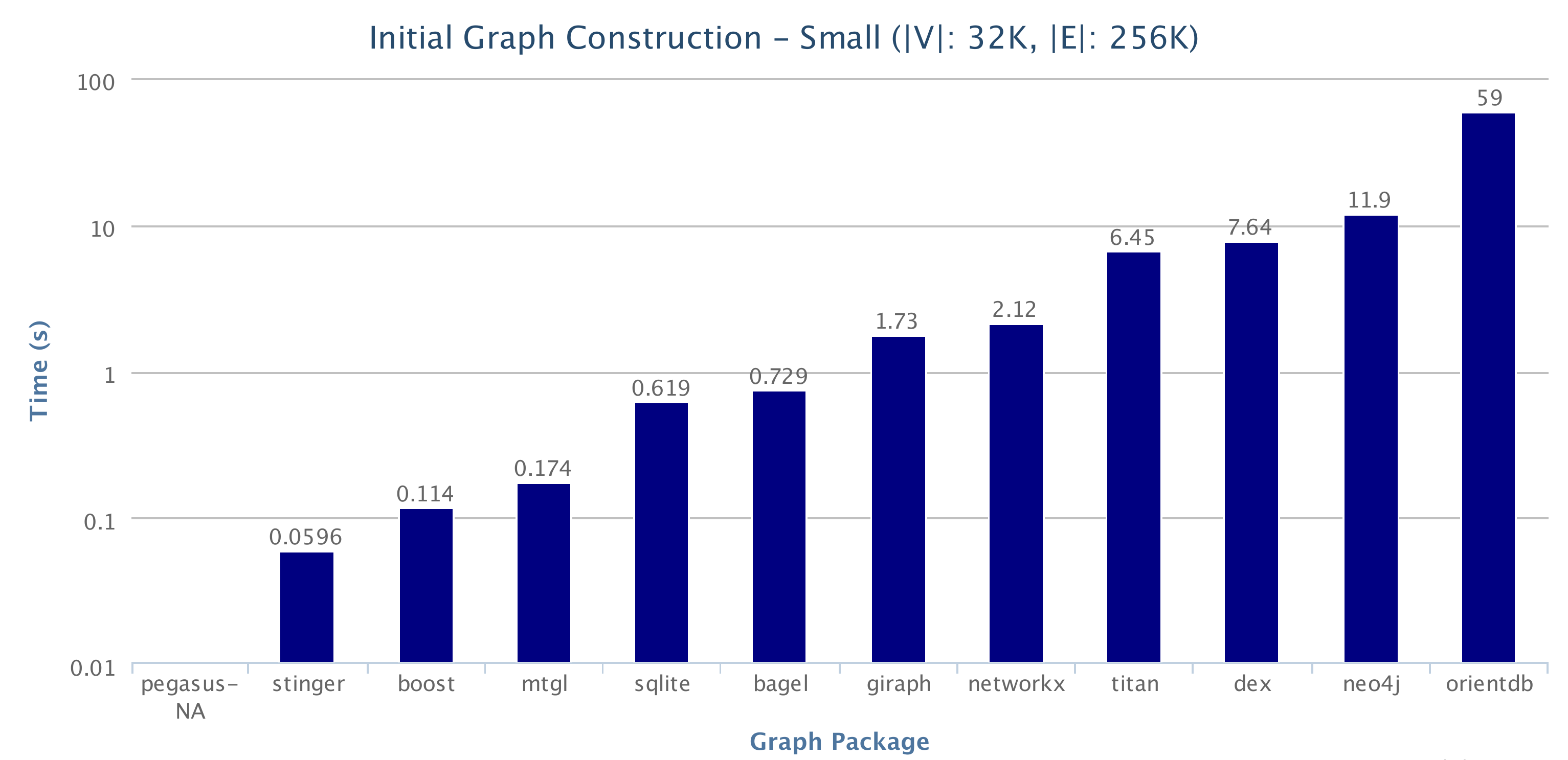}
\caption{The time taken to construct the small graph (32K vertices and 256K undirected edges) for each graph analysis package.}
\end{figure}

\begin{figure}[H]
\centering
\includegraphics[width=0.9\textwidth]{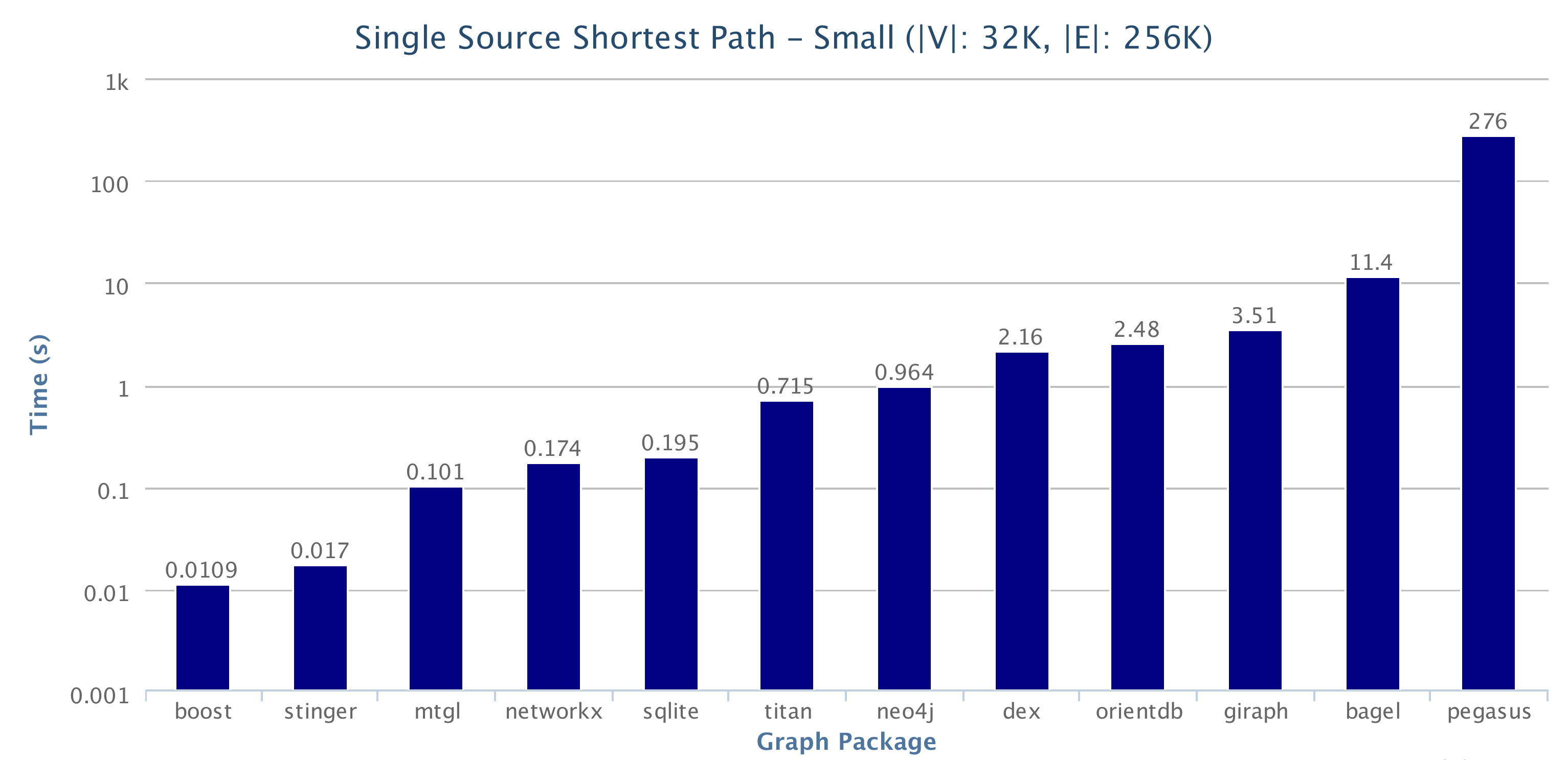}
\caption{The time taken to compute SSSP from vertex 0 in the small graph (32K vertices and 256K undirected edges) for each graph analysis package.}
\end{figure}

\begin{figure}[H]
\centering
\includegraphics[width=0.9\textwidth]{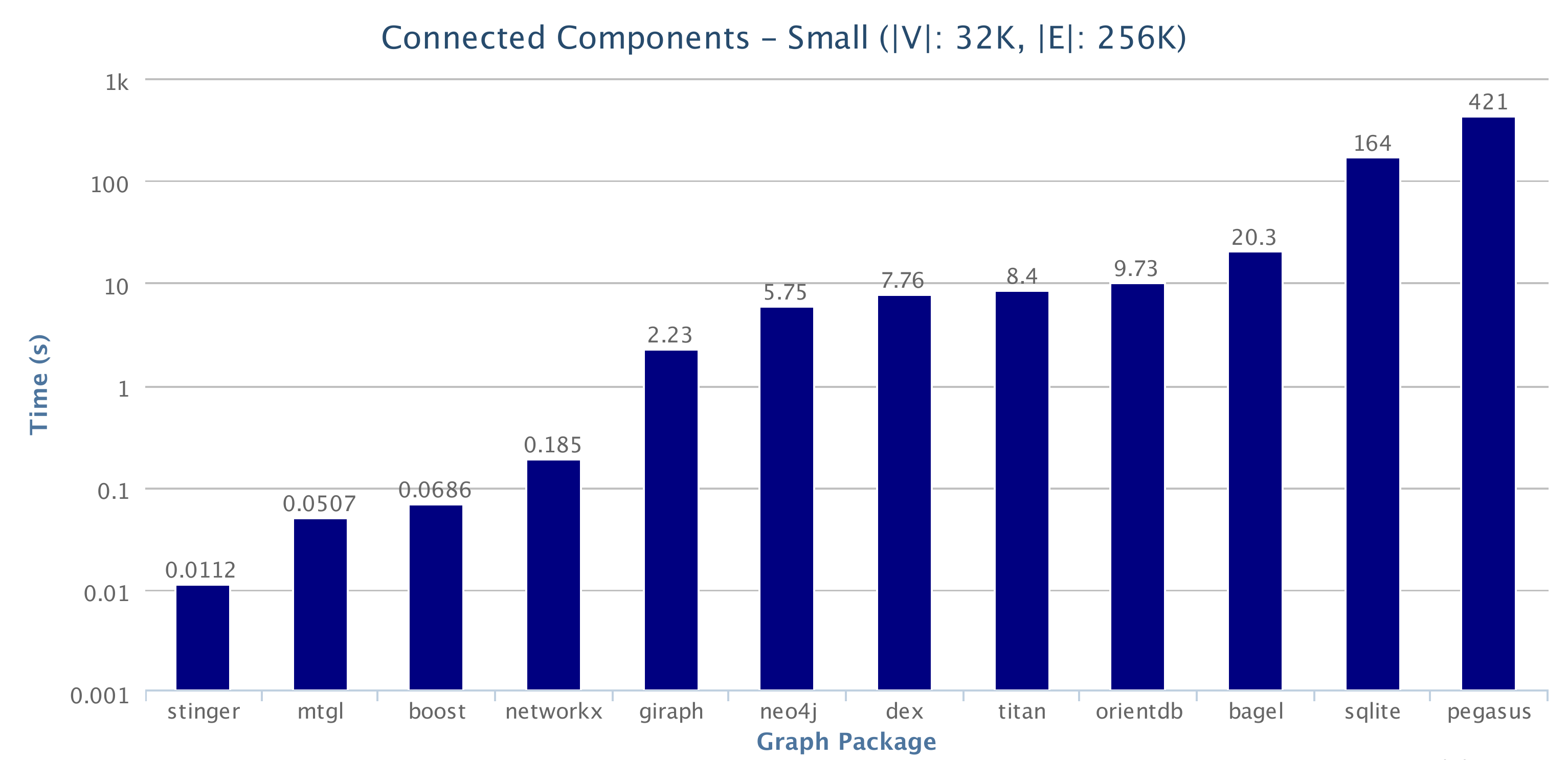}
\caption{The time taken to label the connected components of the small graph (32K vertices and 256K undirected edges) for each graph analysis package.}
\end{figure}

\begin{figure}[H]
\centering
\includegraphics[width=0.9\textwidth]{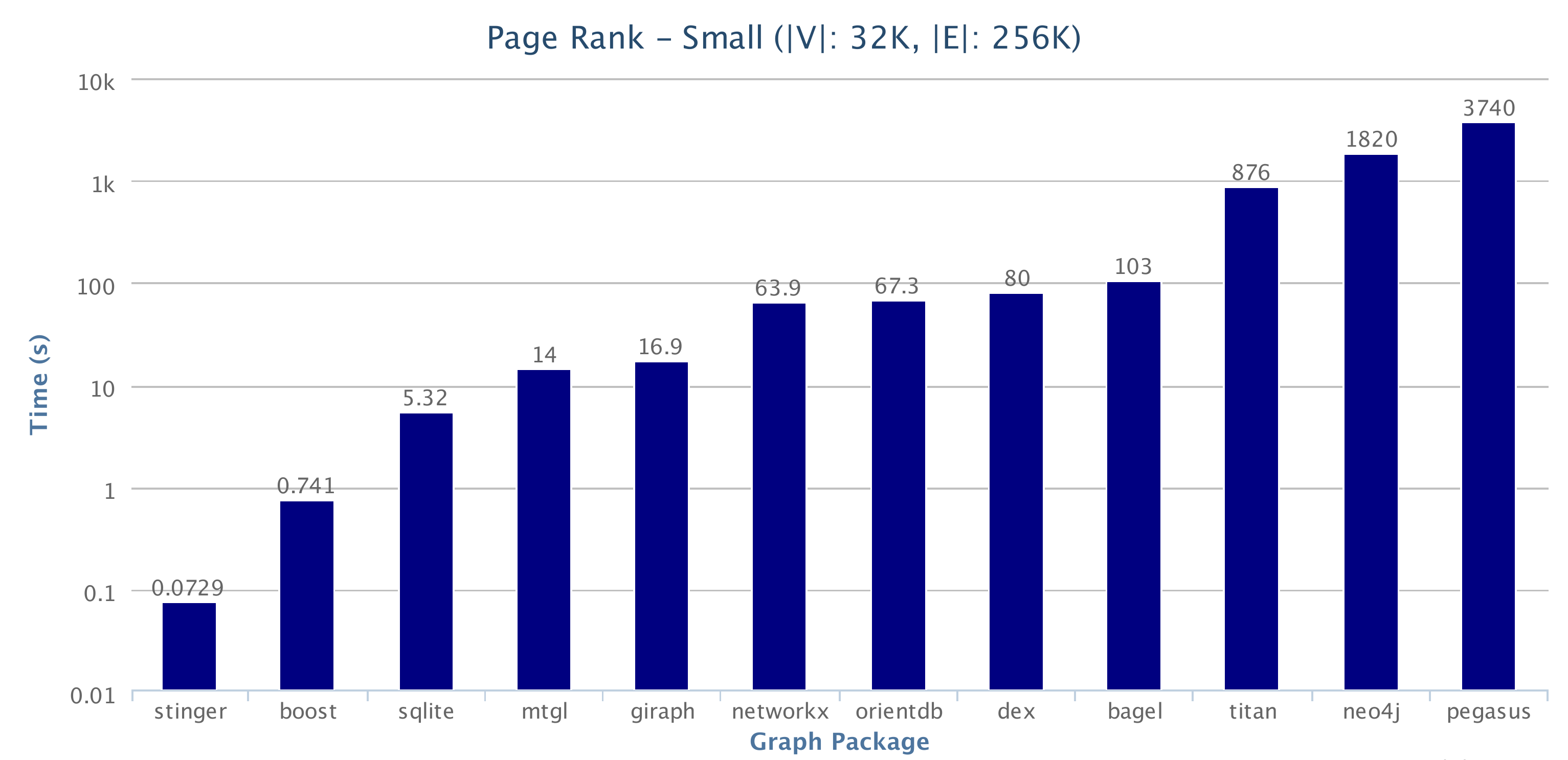}
\caption{The time taken to compute the PageRank of each vertex in the small graph (32K vertices and 256K undirected edges) for each graph analysis package.}
\end{figure}

\begin{figure}[H]
\centering
\includegraphics[width=0.9\textwidth]{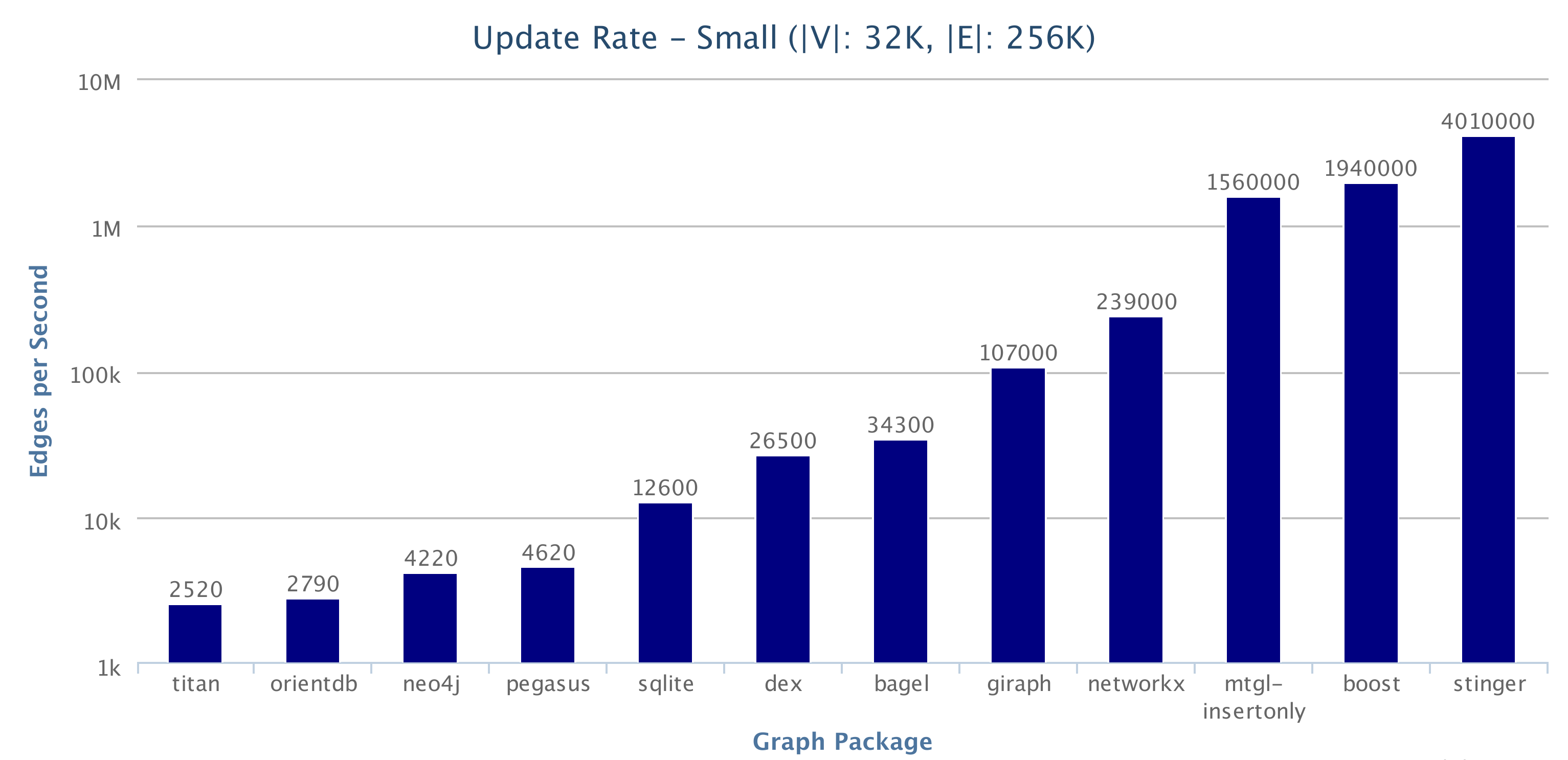}
\caption{The update rate in edges inserted/deleted per second when applying 100,000 edge updates to the small graph (32K vertices and 256K undirected edges) for each graph analysis package.}
\end{figure}

\begin{figure}[H]
\centering
\includegraphics[width=0.9\textwidth]{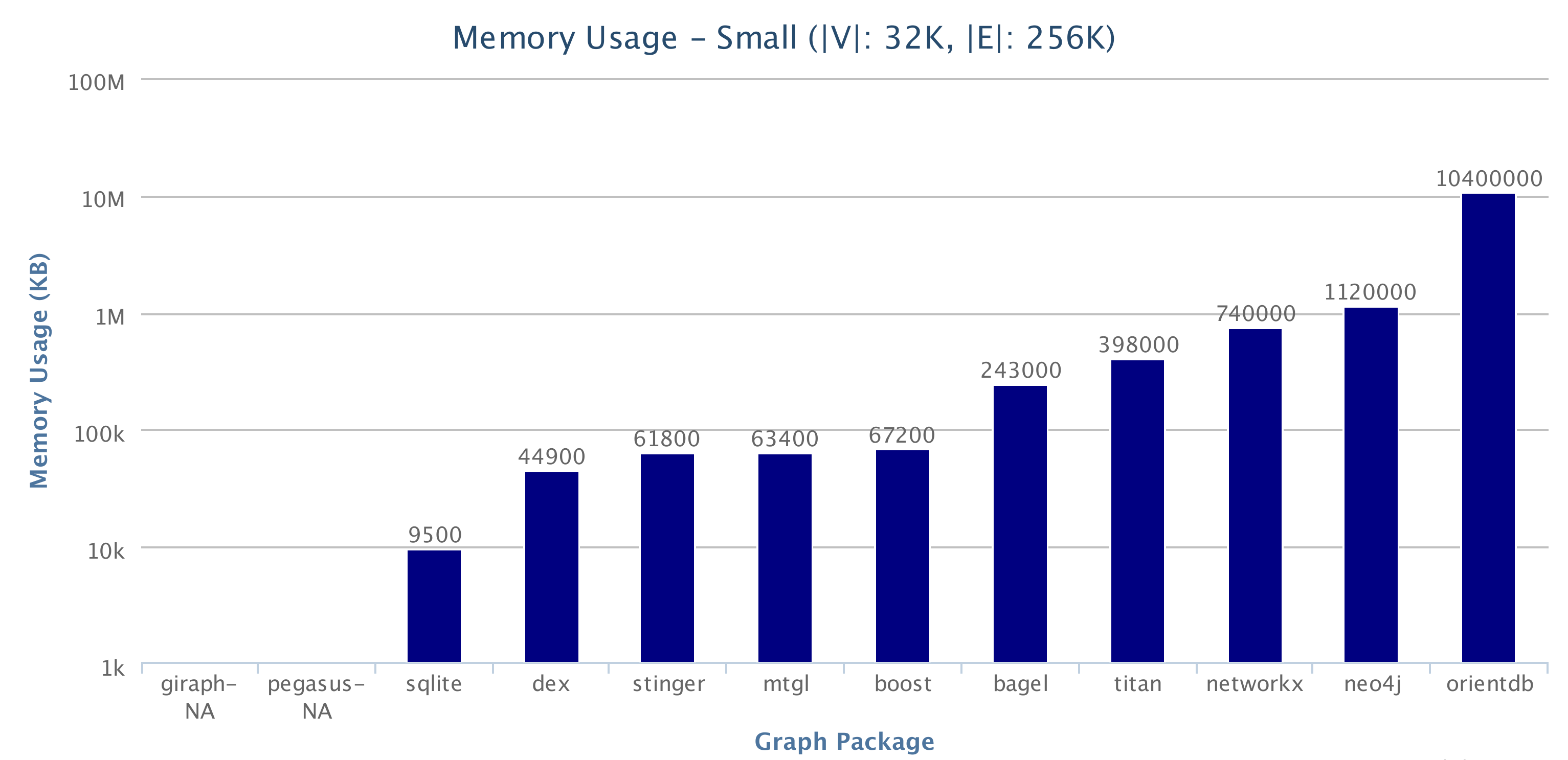}
\caption{The memory occupied by the program after performing all operations on the small graph (32K vertices and 256K undirected edges) for each graph analysis package.}
\end{figure}

\subsection{Medium Graph}

\begin{figure}[H]
\centering
\includegraphics[width=0.9\textwidth]{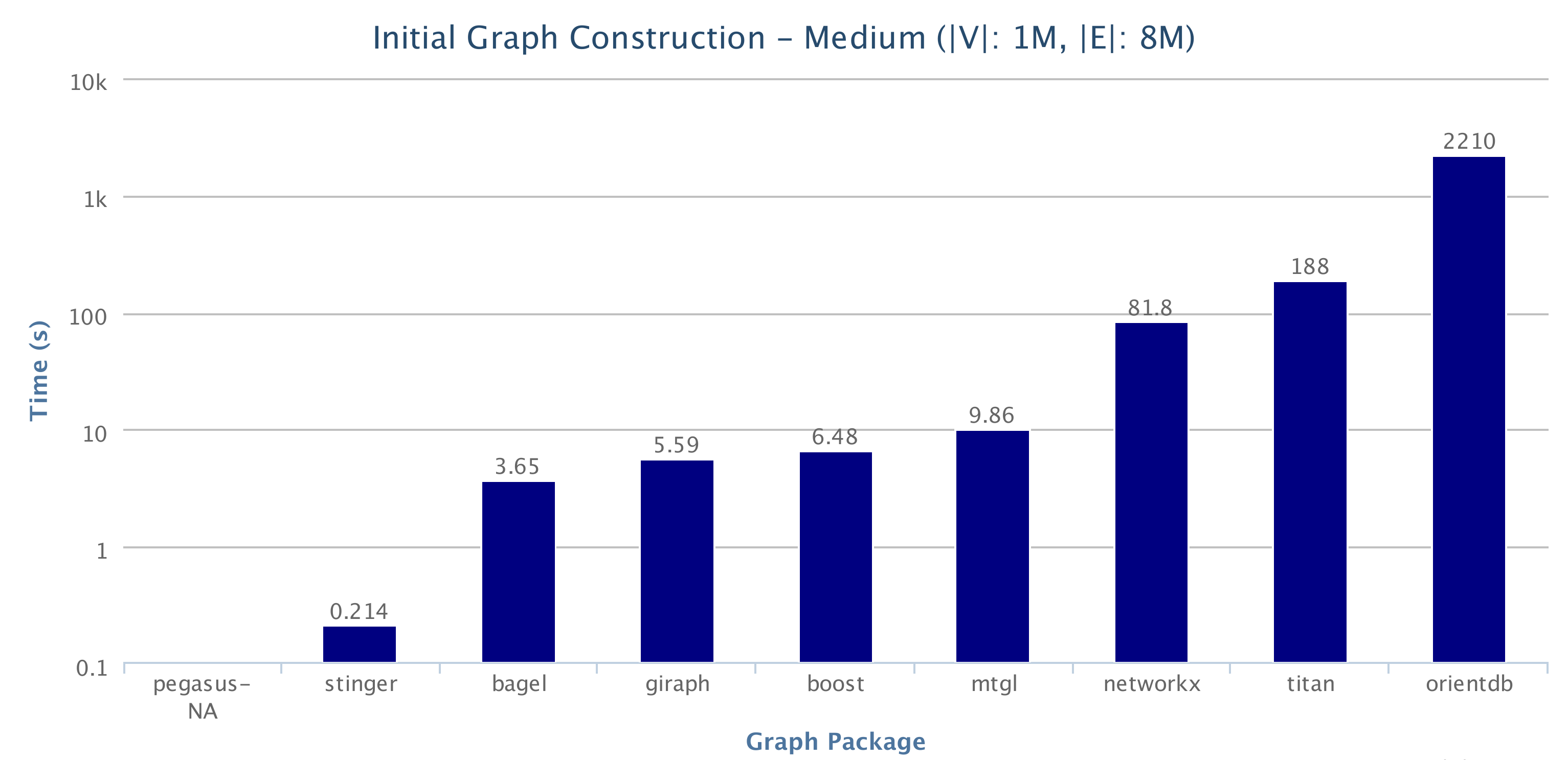}
\caption{The time taken to construct the medium graph (1M vertices and 8M undirected edges) for each graph analysis package.}
\end{figure}

\begin{figure}[H]
\centering
\includegraphics[width=0.9\textwidth]{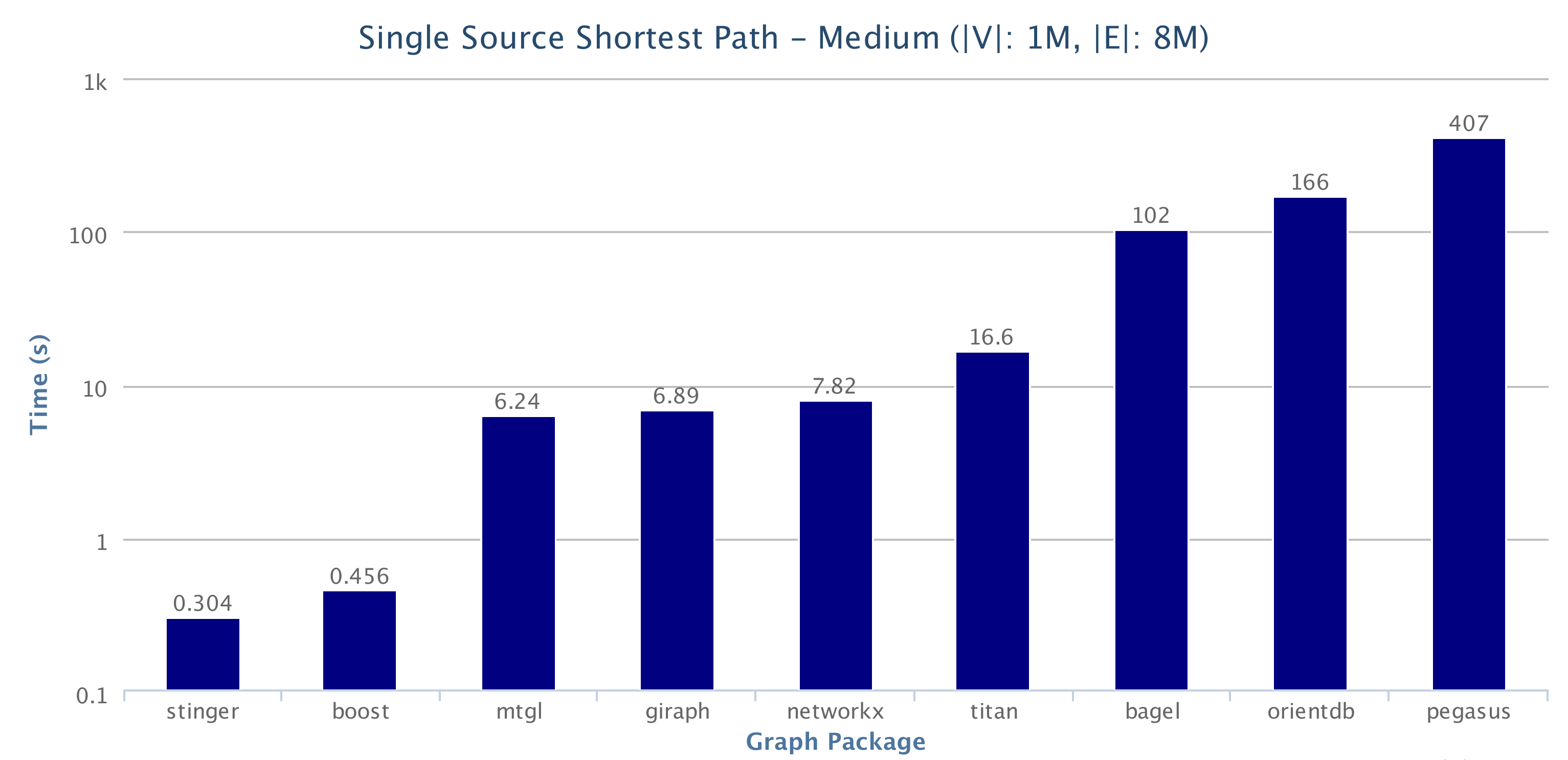}
\caption{The time taken to compute SSSP from vertex 0 in the medium graph (1M vertices and 8M undirected edges) for each graph analysis package.}
\end{figure}

\begin{figure}[H]
\centering
\includegraphics[width=0.9\textwidth]{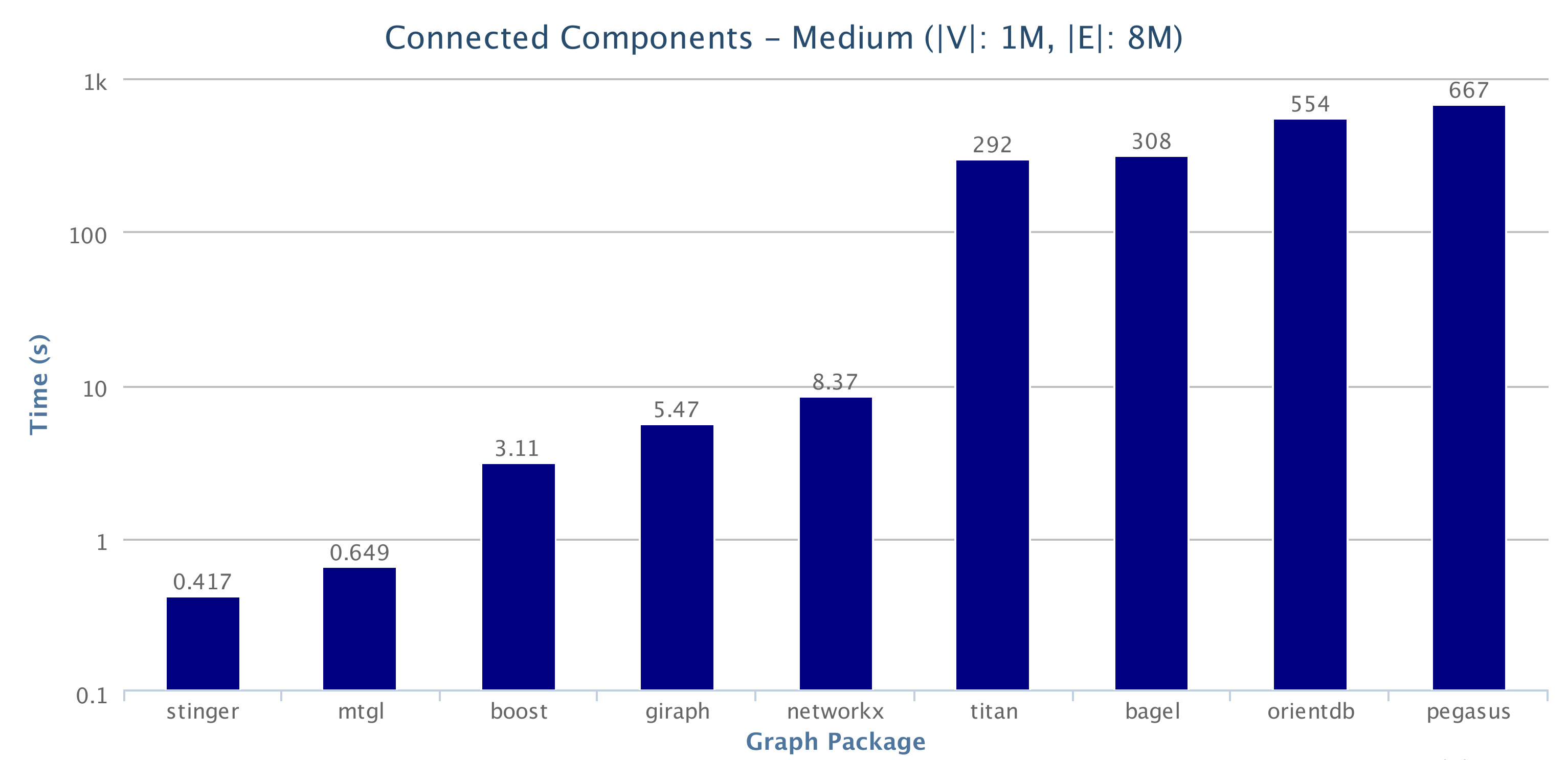}
\caption{The time taken to label the connected components of the medium graph (1M vertices and 8M undirected edges) for each graph analysis package.}
\end{figure}

\begin{figure}[H]
\centering
\includegraphics[width=0.9\textwidth]{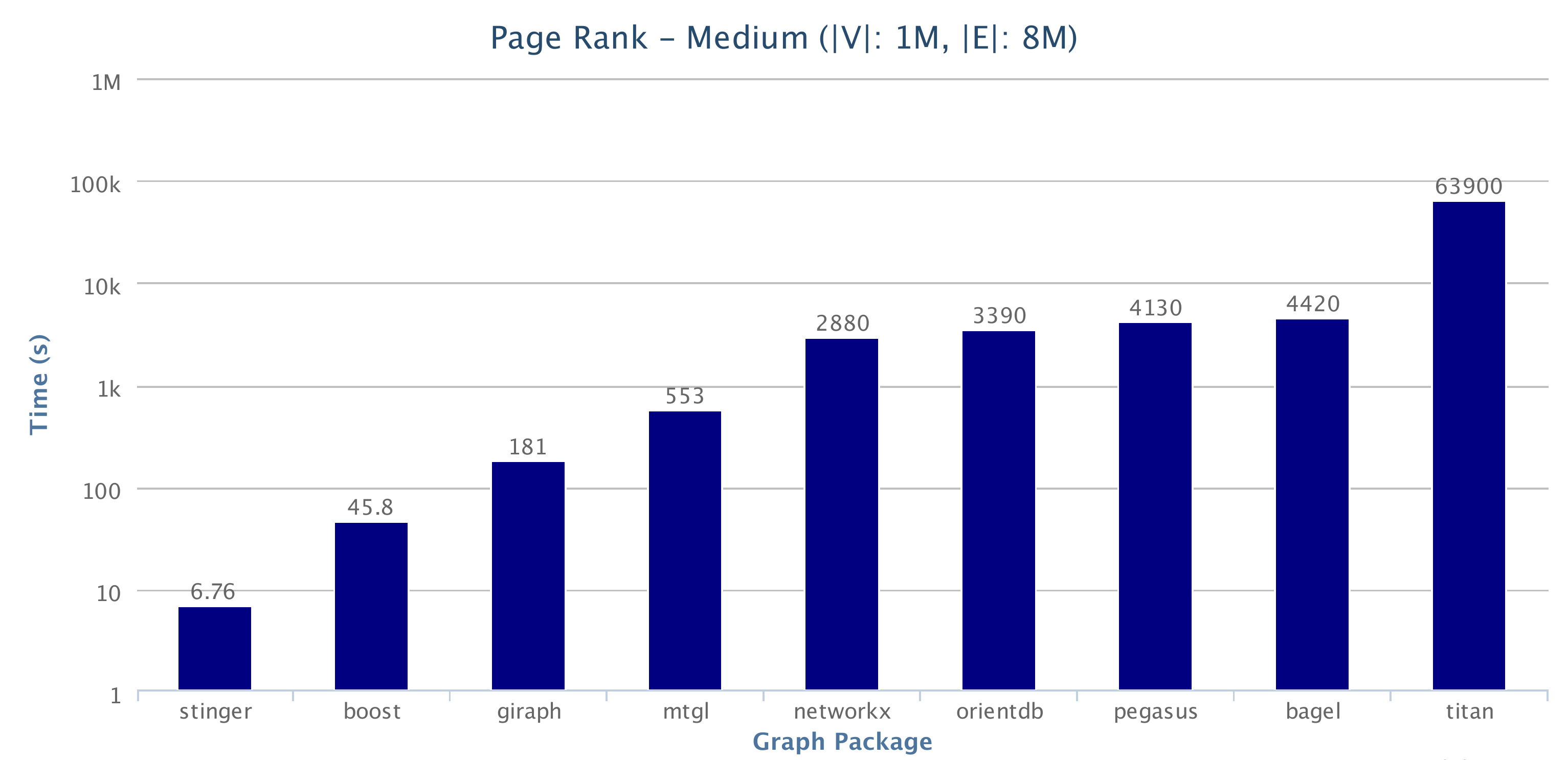}
\caption{The time taken to compute the PageRank of each vertex in the medium graph (1M vertices and 8M undirected edges) for each graph analysis package.}
\end{figure}

\begin{figure}[H]
\centering
\includegraphics[width=0.9\textwidth]{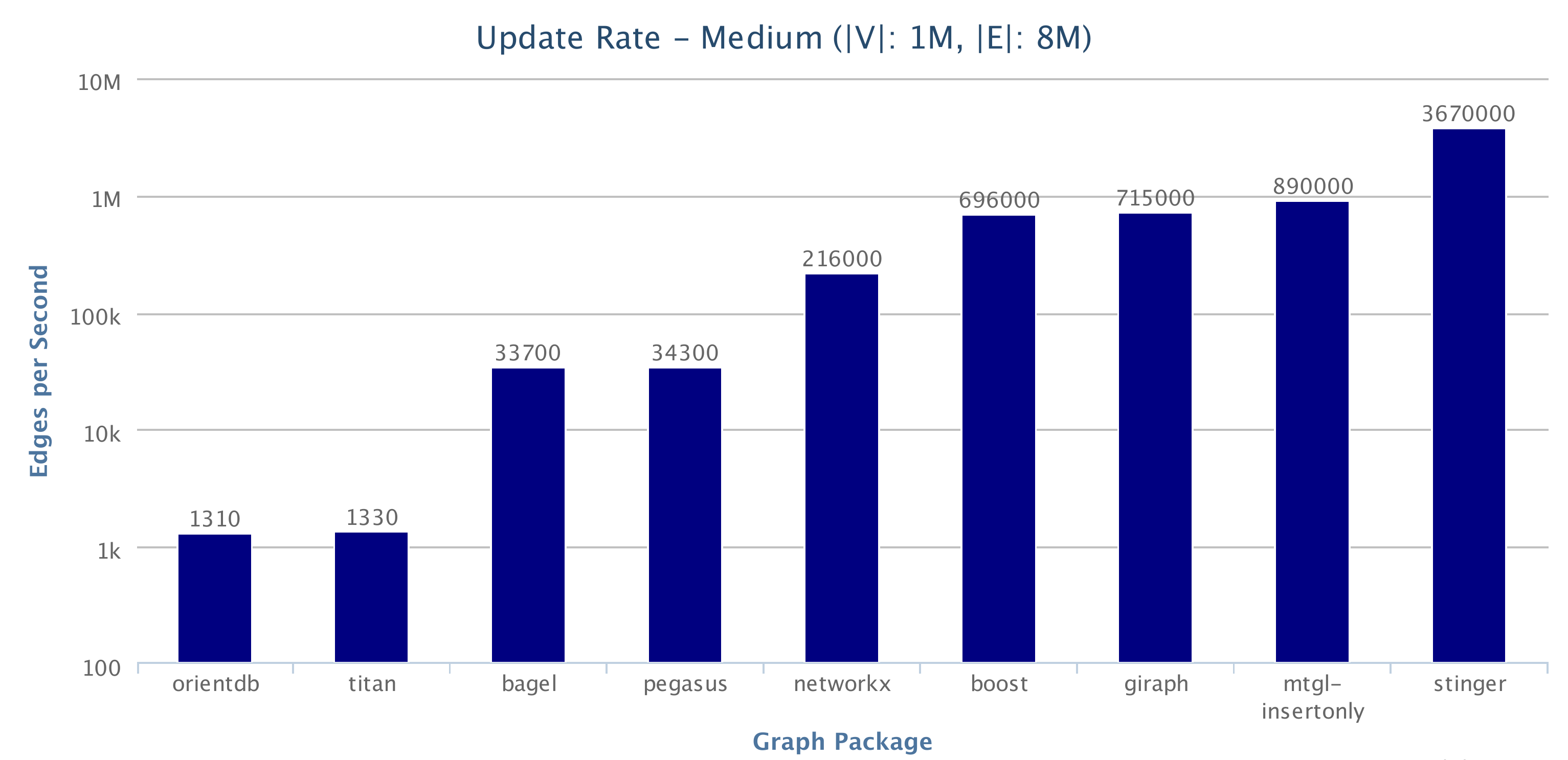}
\caption{The update rate in edges inserted/deleted per second when applying 100,000 edge updates to the medium graph (1M vertices and 8M undirected edges) for each graph analysis package.}
\end{figure}

\begin{figure}[H]
\centering
\includegraphics[width=0.9\textwidth]{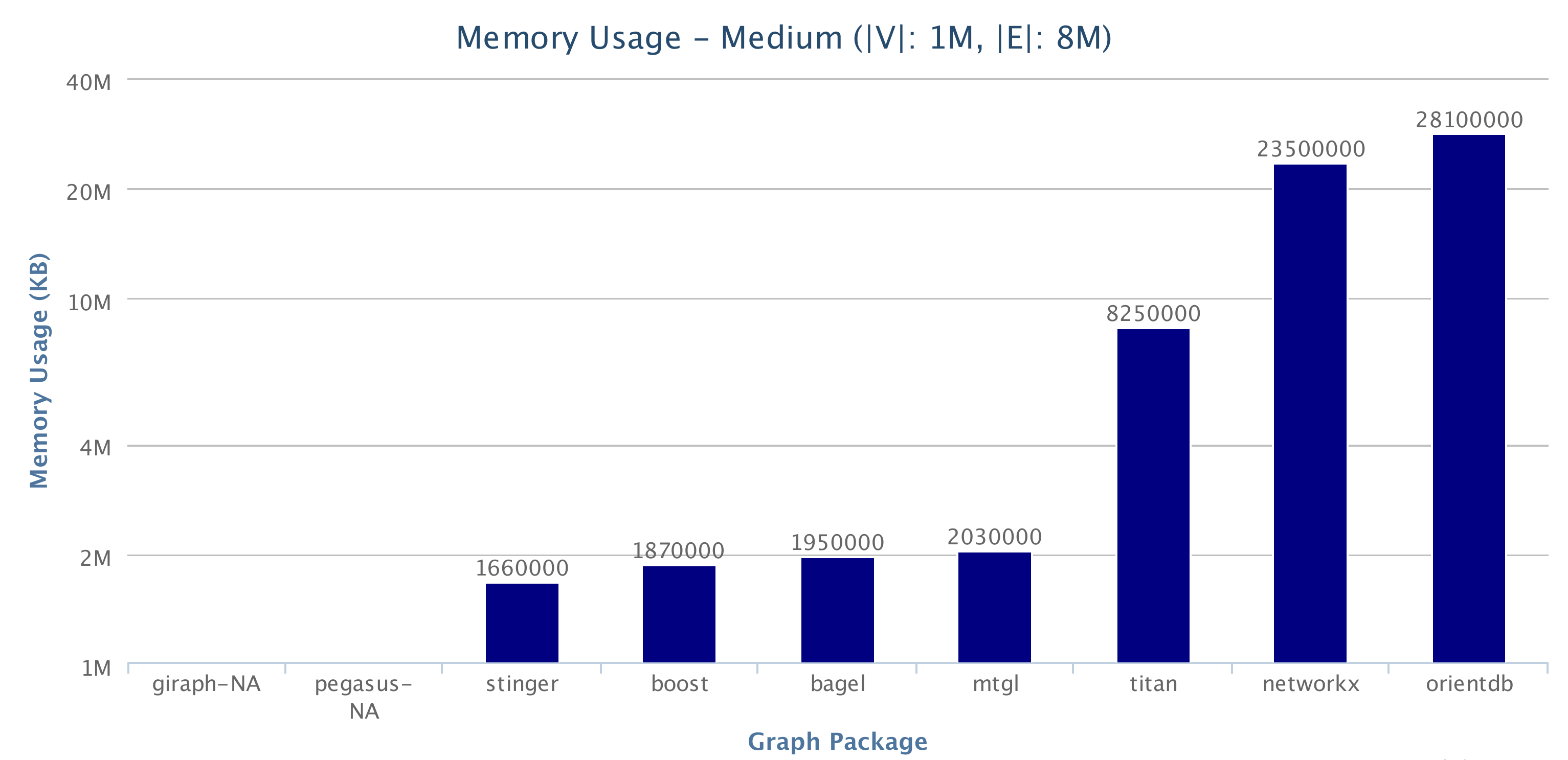}
\caption{The memory occupied by the program after performing all operations on the medium graph (1M vertices and 8M undirected edges) for each graph analysis package.}
\end{figure}

\subsection{Large Graph}

\begin{figure}[H]
\centering
\includegraphics[width=0.9\textwidth]{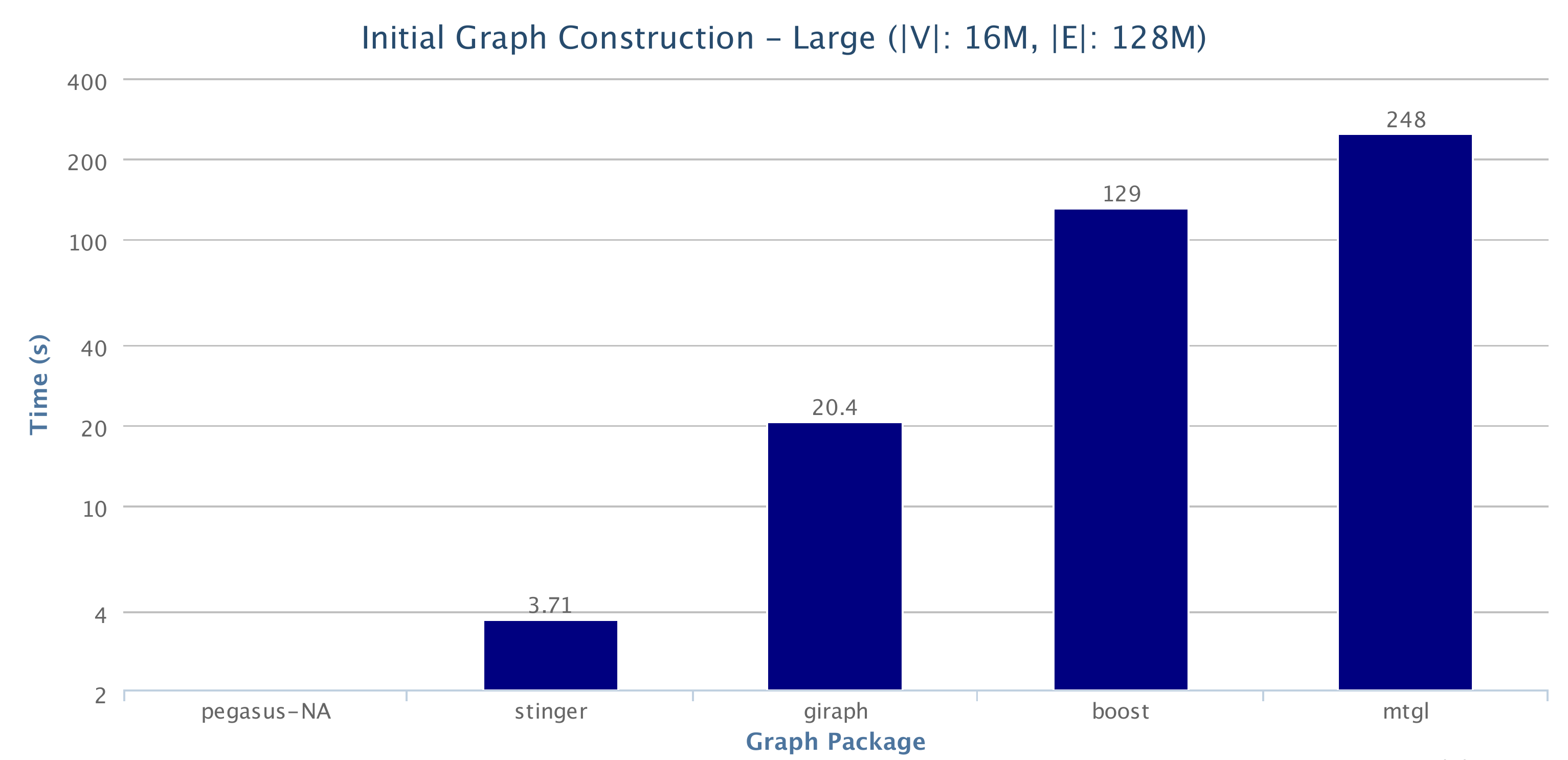}
\caption{The time taken to construct the large graph (16M vertices and 128M undirected edges) for each graph analysis package.}
\end{figure}

\begin{figure}[H]
\centering
\includegraphics[width=0.9\textwidth]{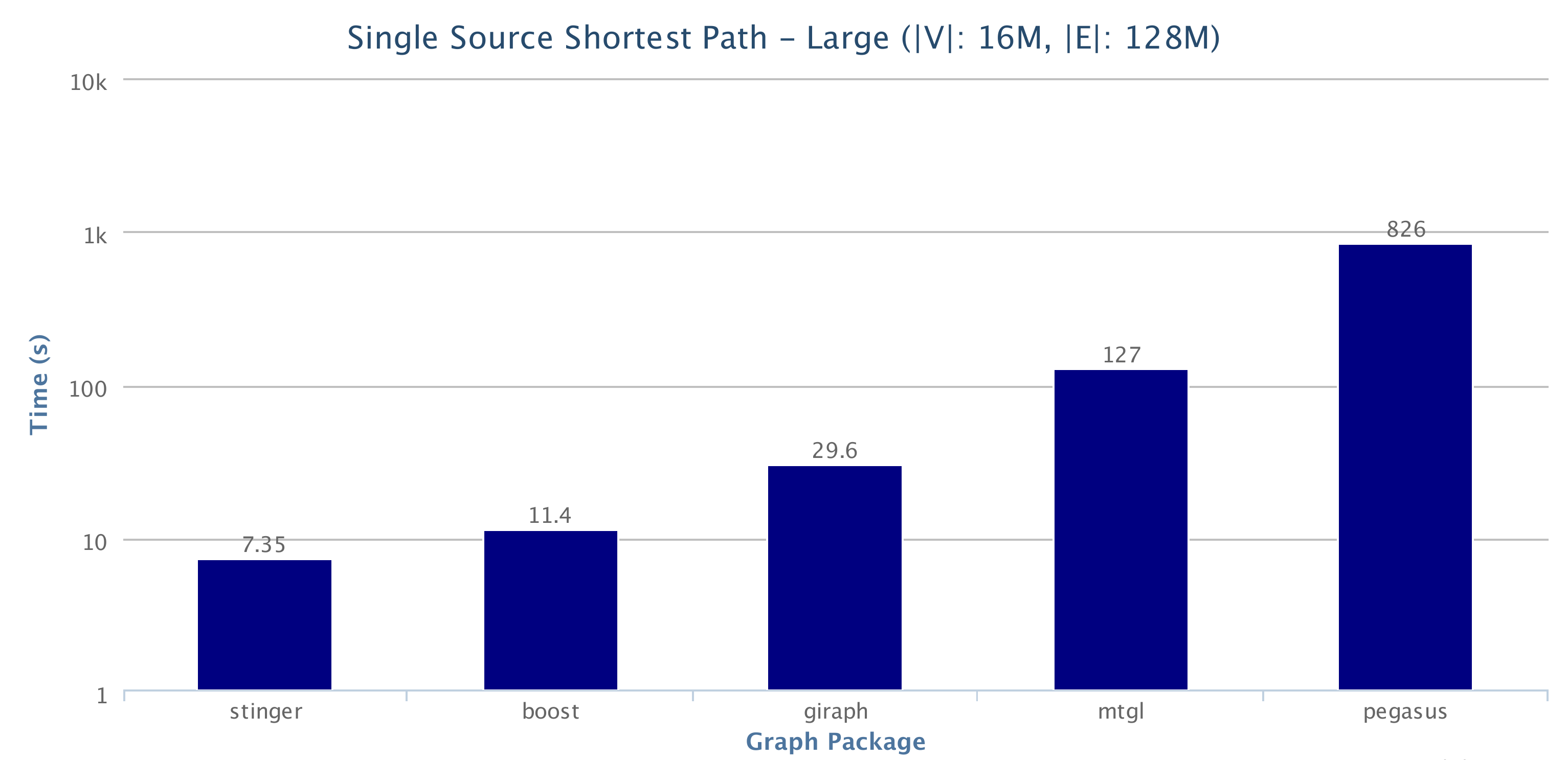}
\caption{The time taken to compute SSSP from vertex 0 in the large graph (16M vertices and 128M undirected edges) for each graph analysis package.}
\end{figure}

\begin{figure}[H]
\centering
\includegraphics[width=0.9\textwidth]{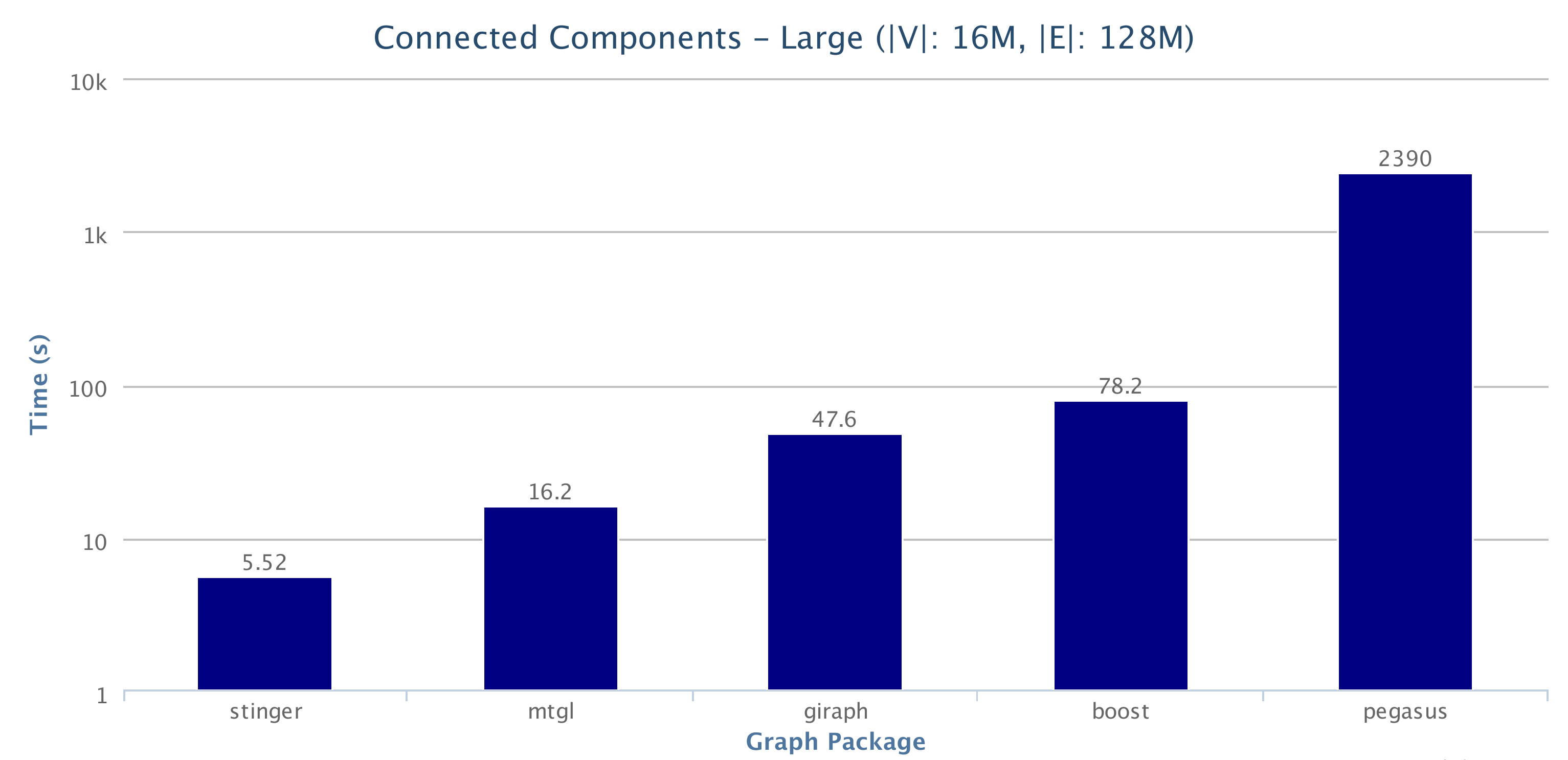}
\caption{The time taken to label the connected components of the large graph (16M vertices and 128M undirected edges) for each graph analysis package.}
\end{figure}

\begin{figure}[H]
\centering
\includegraphics[width=0.9\textwidth]{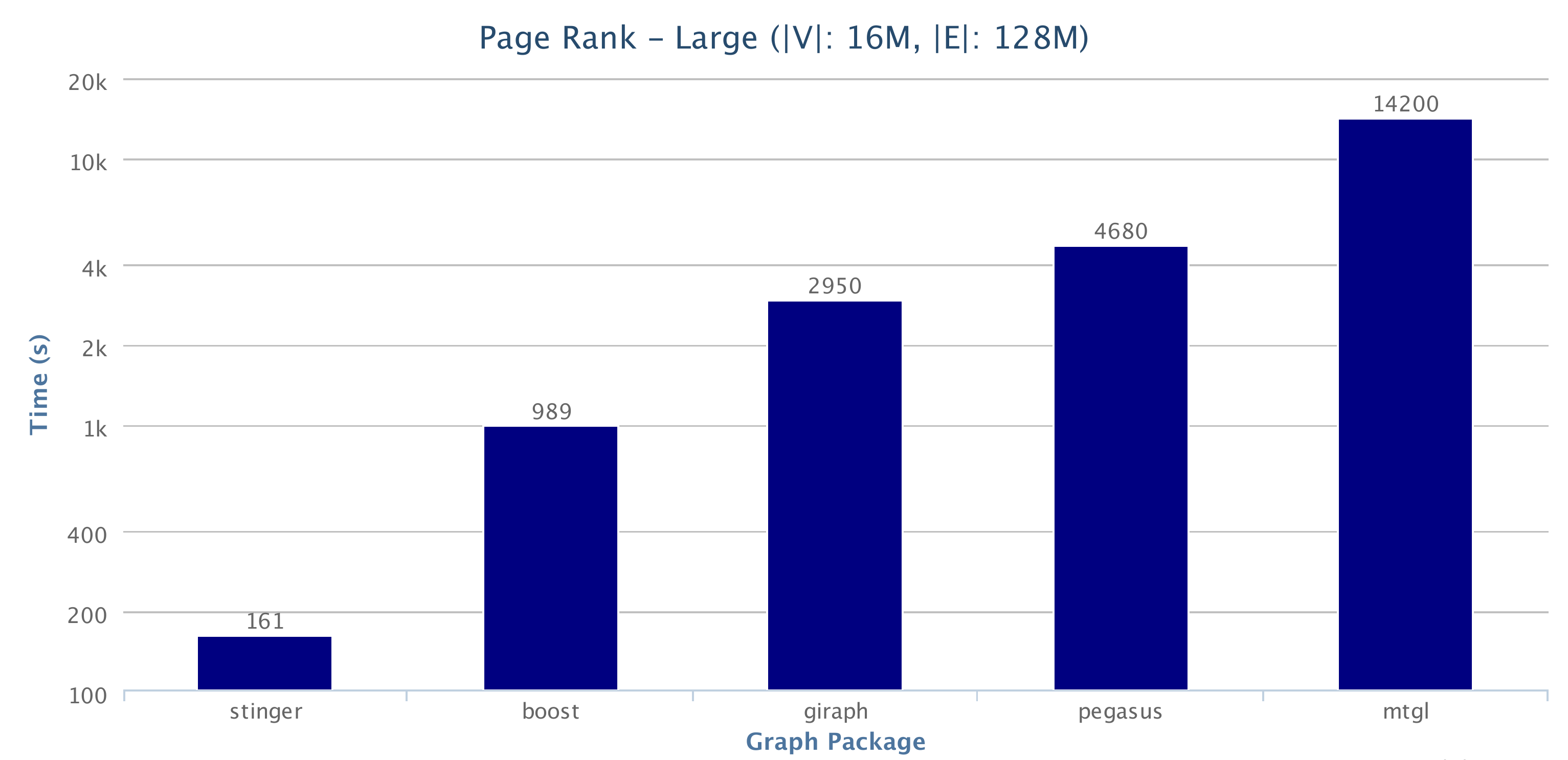}
\caption{The time taken to compute the PageRank of each vertex in the large graph (16M vertices and 128M undirected edges) for each graph analysis package.}
\end{figure}

\begin{figure}[H]
\centering
\includegraphics[width=0.9\textwidth]{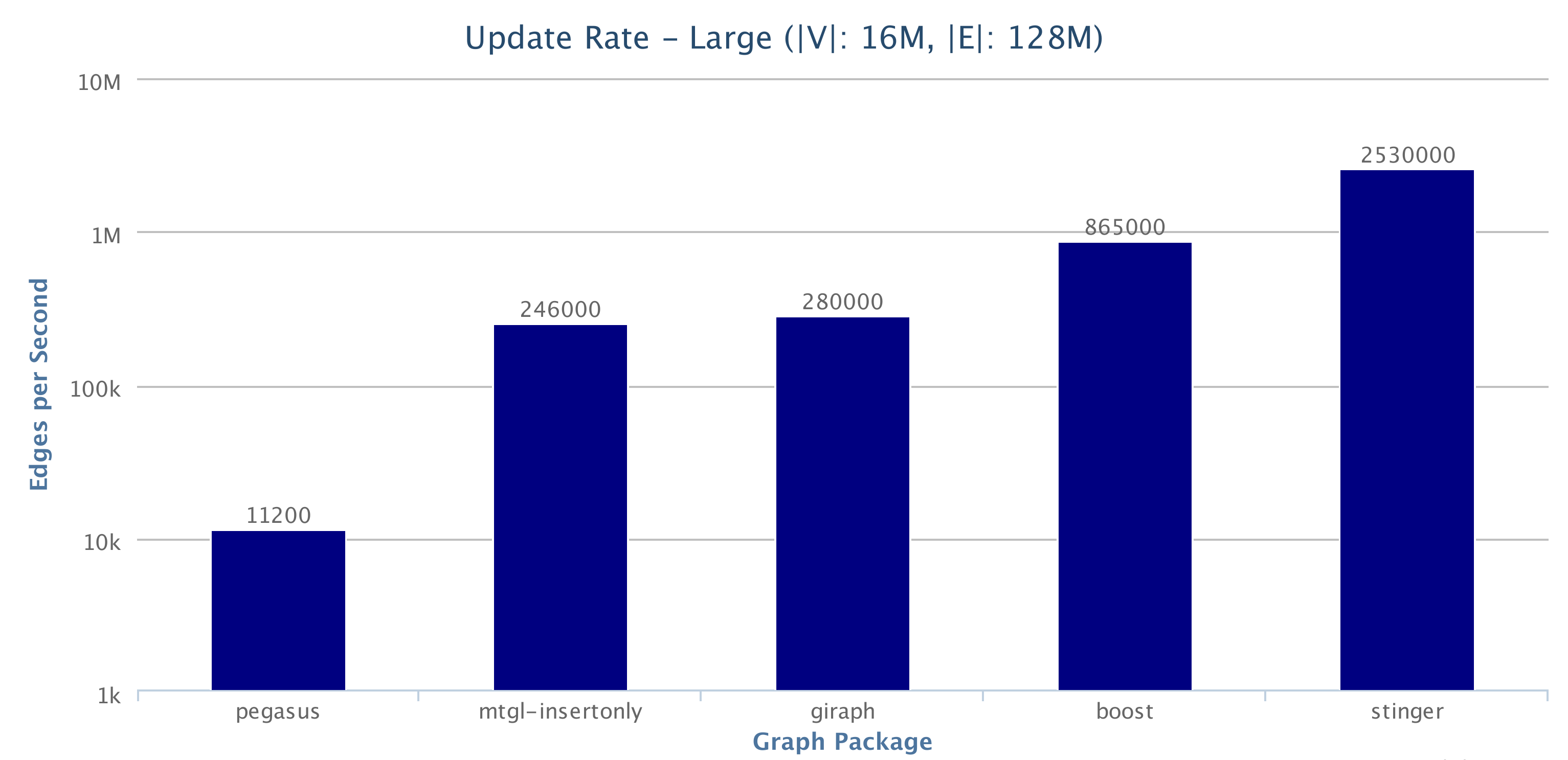}
\caption{The update rate in edges inserted/deleted per second when applying 100,000 edge updates to the large graph (16M vertices and 128M undirected edges) for each graph analysis package.}
\end{figure}

\begin{figure}[H]
\centering
\includegraphics[width=0.9\textwidth]{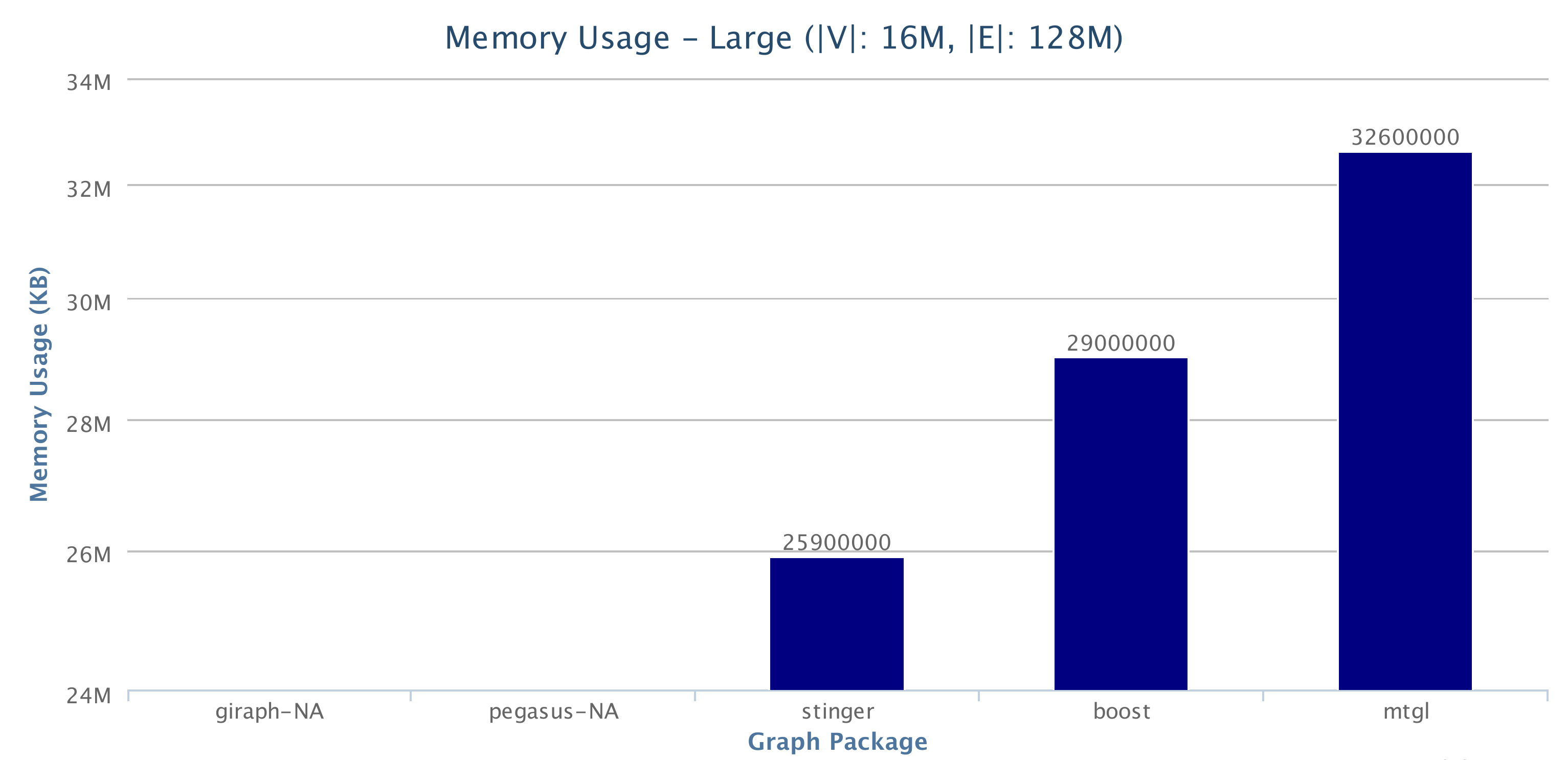}
\caption{The memory occupied by the program after performing all operations on the large graph (16M vertices and 128M undirected edges) for each graph analysis package.}
\end{figure}

\end{document}